\documentclass[prd, twocolumn, lengthcheck, superscriptaddress]{revtex4-1}
\usepackage{graphicx}
\usepackage{url}
\usepackage{amssymb} %,amsmath,longtable,rotating,color,units,wasysym,subfigure,epsfig,listings,bm}
\usepackage{hyperref}
\usepackage{cleveref}
\usepackage{times}
\usepackage{amsmath}
\usepackage{graphicx}
\usepackage{epsfig, units}
\DeclareGraphicsExtensions{.pdf,.eps,.png,.jpg}
\usepackage{acronym}
\usepackage{color}
\usepackage{longtable}
\usepackage[caption=false]{subfig}
\usepackage{graphicx,url,amssymb,longtable,rotating,color,units,wasysym,epsfig,listings,bm}
\usepackage[section]{placeins}
\usepackage{algorithm}
\usepackage{algorithmic}
\usepackage{afterpage}

\newcommand{\subfigimg}[3][,]{%
  \setbox1=\hbox{\includegraphics[#1]{#3}}% Store image in box
  \leavevmode\rlap{\usebox1}% Print image
  \rlap{\hspace*{10pt}\raisebox{\dimexpr\ht1-2\baselineskip}{#2}}% Print label
  \phantom{\usebox1}% Insert appropriate spcing
}

\begin{document}

\title{Sensitivity study using machine learning algorithms on simulated r-mode gravitational wave signals from newborn neutron stars}

\author{Antonis Mytidis}
\affiliation{Department of Physics, University of Florida, 2001 Museum Road, Gainesville, FL 32611-8440, USA} 

\author{Athanasios Aris Panagopoulos}
\affiliation{Department of Computer Science, California State University Fresno, 5241 N. Maple Ave, Fresno, CA 93740, USA}

\author{Orestis P. Panagopoulos}
\affiliation{Department of Computer Information Systems, California State University Stanislaus, One University Circle, Turlock, CA 95382, USA}

\author{Andrew Miller}
\affiliation{Department of Physics, University of Florida, 2001 Museum Road, Gainesville, FL 32611-8440, USA}
\affiliation{INFN, Sezione di Roma, I-00185 Roma, Italy}
\affiliation{Universit\`a di Roma La Sapienza, I-00185 Roma, Italy}

\author{Bernard Whiting}
\affiliation{Department of Physics, University of Florida, 2001 Museum Road, Gainesville, FL 32611-8440, USA}

\begin{abstract}
This is a follow-up sensitivity study on r-mode gravitational wave signals from newborn neutron stars illustrating the applicability of machine learning 
algorithms for the detection of long-lived gravitational-wave transients. In this sensitivity study we examine three machine learning algorithms (MLAs):
artificial neural networks (ANNs), support vector machines (SVMs) and constrained subspace classifiers (CSCs). The objective of this study is to compare the 
detection efficiencies that MLAs can achieve to the efficiency of the conventional (seedless clustering) detection algorithm discussed in an earlier paper. 
Comparisons are made using 2 distinct r-mode waveforms. For the training of the MLAs we assumed that some information about the distance to the source is 
given so that the training was performed over distance ranges not wider than half an order of magnitude. The results of this study suggest that we can use 
the machine learning algorithms as part of an investigative stage in the pipeline that would be able to provide very fast and solid triggers for further, 
and more intense, investigation.\\ 
\end{abstract}

\maketitle

\section{Introduction}
\label{sec:intro}

In the late 1990s, the r-mode quasi-toroidal pulsations of a neutron star became very promising for generating strong gravitational-wave signals due to the 
Chandrasekhar-Friedman-Schutz (CFS) instability they exhibit \citep{SOLTWOPRBL,LAGRANGEPERT,SECINSTROTNS}. R-modes of any harmonic, frequency and amplitude are 
subject to this instability at any angular velocity of the star \citep{NCRM1997,JFSM1998}. Therefore, even the smallest toroidal perturbations in the velocity of 
the neutron star mass currents will keep increasing in amplitude. In the absence of a saturation mechanism these small perturbations could eventually reach energy 
values of the order of the rotational energy of the neutron star. 

In considering the saturation amplitude, $\alpha$, its normalization is such that values of order 1 carry energy of the same order of magnitude as the total rotational 
energy of the neutron star. Some authors have introduced damping mechanisms that can cause saturation at r-mode oscillation amplitudes of order $10^{-4}-10^{-2}$ 
dimensionless units \citep{SDNS2009}, while others have introduced mechanisms that cause saturation at amplitudes equal to or larger than $10^{-1}$ \citep{ALFORD}. 
Some of the factors that can affect the order of $\alpha$ are: the equation of state (EOS) of the matter in the center of the neutron 
star \citep{rmodespaper}, the magnitude of the magnetic fields on the neutron star \citep{HOLAI, HKTOP}, the coupling of the r-modes with other inertial 
modes \citep{NONLINCOUPLROTNS} and magneto-hydrodynamic coupling to the stellar magnetic field \citep{rezzolla2000r}. Therefore, an r-mode detection and a subsequent 
estimation of the saturation amplitude will impact all of the above theories depending on the order of magnitude of $\alpha$ they predict.

The physical significance of an r-mode gravitational-wave detection has been extensively studied over the past 15 years \citep{GWHYNS, GRINSTHYNS, RMODEINST}. 
Theoretical studies suggest that (assuming the r-mode oscillation amplitude grows sufficiently large) r-mode gravitational radiation (primarily in the $m = 2$ harmonic)
could carry away most of the angular momentum of a rapidly rotating newborn neutron star. Therefore, an r-mode detection would also (i) provide explanation of the low rotational frequencies of the observed neutron 
stars when compared to their possible rotational frequencies at birth, (ii) set constraints on the equation of state of the matter in the core of the neutron star and 
(iii) set upper bounds on $\alpha$ and settle the debate about the magnitude of the saturation amplitude of the r-mode oscillations on neutron stars \citep{SPINOTT}.

In a previous study we argued that the most promising r-mode gravitational-wave sources are newborn neutron stars \citep{rmodespaper}. In subsection \ref{model} (equation \eqref{eq.3aa})
we show that, due to their high angular velocities newborn neutron stars will emit the most powerful r-mode gravitational radiation among all other possible sources. 
Therefore, the design of an r-mode search from newborn neutron stars depends on an electromagnetic trigger from a supernova (type-I or type-II) event. The distance to the 
r-mode source is needed to extract any information about the magnitude of $\alpha$ because an r-mode detection can only give an estimate for the ratio $\alpha /d$, as shown 
in section \ref{Design}, equation \eqref{eq.3a}. Distances to type-I supernova can be calculated using the standard candle method with an error between $\unit[5-10]{\%}$ \citep{TYPE1DIST}.
Distances to type-II supernovae can be calculated using the expanding photosphere method giving an error of $\unit[10-15]{\%}$ \citep{photospheric,TYPEIIDIST}.  

The results of our previous sensitivity study showed that advanced LIGO (aLIGO) can be sensitive to r-mode signals from newborn neutron stars only within our local group 
of galaxies. Since distances to galaxies in our local group are already known a supernova event within our local group would automatically give information about the 
distance to the hypothetical r-mode gravitational radiation source. The latest supernova event in our local group (SN2014J) occurred in January of 2014 in the galaxy Messier 
82 (M82) in the nearby group of galaxies M81 and it was a type-I supernova \citep{SN2014J}. This galaxy is at a distance of $\unit[3.5]{Mpc}$ from the Earth. This is a factor 
of 3 further than our previous sensitivity study showed that aLIGO can be sensitive at. At that distance the supernova event rate is only 3-6 per century \citep{SNRATE,MLG} 
and the best we can do in order to be ready for the next event is to increase the sensitivity of our algorithms or apply a new class of more efficient algorithms.

In this study we investigated the applicability of machine learning algorithms (MLAs) as decision makers (signal or not) for the detection of r-mode gravitational waves.  
This study was performed by integrating the MLAs in the stochastic pipeline \citep{STAMPPAPER, advancedLIGO_35_6, LIGO_93_042005}. The objective of this pipeline is to explore 
the possibility of sources of {\it long-lived} gravitational-wave transients lasting from many seconds to weeks. Searches for long-lived gravitational-wave transients 
have a strong scientific motivation. This is a cross-correlation-based analysis pipeline, which was formed to bridge the gap between short ${\cal O}(s)$ burst analyses 
and stochastic analyses (in which the signal is assumed to persist through the duration of the data-taking run). The pipeline is framed as a pattern recognition problem.

The investigation we present in this paper is a preliminary one with the target to initiate further research in this field. The purpose of this paper is to provide some 
insight into how we can harness the power of MLAs and use them for the r-mode gravitational-wave searches. Additionally, the methodology we followed here may also be used 
for a broader investigation of the applicability of MLAs for the detection of other long-lived gravitational-wave transients. The aim of this paper is not to demonstrate 
how we can use MLAs in order to make a detection announcement. Instead, the aim is to perform a preliminary investigation on how we can use raw data taken by the LIGO detectors, 
pre-process it and feed it into three separate MLAs. The ultimate target is to use MLAs in a way that we can facilitate the searches for long-lived gravitational-wave 
transients in the stochastic pipeline.

\subsection{R-mode model}
\label{model}

The r-mode gravitational-wave model we used in our present study as well as in our previous work \citep{rmodespaper}, is based on the \text{Owen et al.\,\,\,}'98 model.
Though very simplistic, this model is still a very good approximation for the early stages of the neutron star spin-down \citep{GWHYNS}. More complicated numerical methods 
have shown that an r-mode saturation amplitude $\alpha=10^{-2}$ can result in a spin-down whose energy loss can be detected as gravitational radiation by aLIGO \citep{SDNS2009}. 
When this saturation amplitude is used in the '98 model, we see that there is a good agreement in the angular velocity evolution of the neutron star up to several months after 
the start of the neutron star spin-down \citep{rmodespaper}. The evolution of the gravitational-wave frequency emitted by the neutron star in the Owen et al. '98 model is described by
\begin{equation}
\label{eq.1a}
f(t) = \frac{1}{ \left ( f_o^{-6} + \mu t \right )^{\frac{1}{6}} }
\end{equation}

\noindent
where $\mu$ is an EOS dependent parameter \citep{rmodespaper}. For a polytropic EOS this parameter is expressed as a function of $\alpha$ as follows  
\begin{equation}
\label{eq.2a}
  \mu= 1.1 \times 10^{-20} |\alpha|^2 \frac{\text{s}^{-1}}{\text{Hz}^6}.
\end{equation}

\noindent
For the same model and the same EOS the gravitational radiation power is given by
\begin{equation}
\label{eq.3aa}
\dot{E} \approx 3.5 \times 10^{19} f^8 | \alpha |^2 \,\ \mbox{W}.
\end{equation}

\noindent
This model depends on two parameters: the (dimensionless) saturation amplitude, $\alpha$, of the r-mode oscillations and the initial gravitational 
wave spindown frequency $f_o$. The theoretical predictions for the values of these parameters were discussed extensively in our previous paper. 
The values we considered for $\alpha$ lie in the range of $10^{-3}- 10^{-1}$ while the values we considered for $f_o$ lie in the interval of 
$\unit[600-1600]{Hz}$. Due to the wide range within which the values of these parameters lie, we cannot effectively use a matched filtering algorithm. Instead, 
we have to develop techniques that could detect all possible distinct waveforms. 

\subsection{Previous work}
\label{previous}

In our previous paper, a seedless clustering (SC) algorithm was used \citep{stochtrack}. This seedless clustering
algorithm integrates the signal-to-noise ratio (SNR) of pixels along predetermined monotonic curves (“clusters”) with arbitrary
start and stop times and the constraint that there is a minimum total duration. Then the algorithm
performs the weighted sum of the pixel-SNR values along each curve to calculate the cluster SNR. After
repeating this T-many times (where T is a free parameter of the algorithm), the algorithm records the
largest cluster-SNR value. The algorithm records the largest cluster-SNR value for each one of the ft-maps
that goes through the pipeline.
This method is not dependent on any knowledge of the signal and it can be applied generically 
to any long-lived gravitational-wave transients. In particular, it is unable to discriminate between r-modes and other possible gravitational wave 
sources. Knowledge of the r-mode signal can be used to make minor modifications in the clustering algorithm, however, there was not much hope for a 
dramatic improvement in the efficiency. Nevertheless, we were able to recover signals of magnitude 5 times weaker than the noise. 

\begin{figure}
\includegraphics[width=1.05 \linewidth]{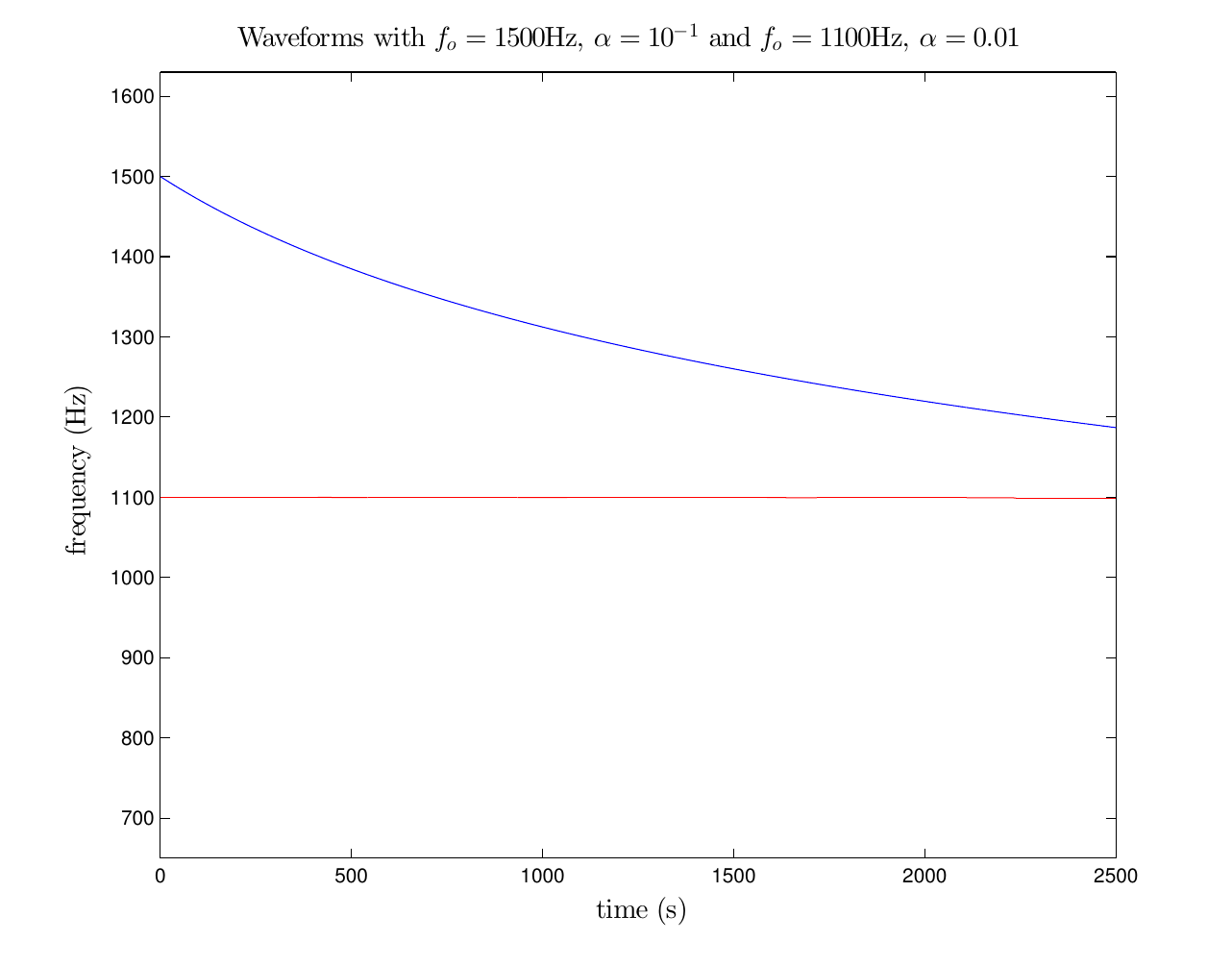}
\caption{The $(f_o=\unit[1500]{Hz}, \alpha=0.1)$ waveform is the most powerful waveform considered in our sensitivity studies both for the clustering and the MLAs. 
         The second waveform we chose has an amplitude 25 times smaller than the first one. This waveform has parameters $(f_o=\unit[1100]{Hz}, \alpha=0.01)$ and is 
         approximately monochromatic for the durations our sensitivity studies were designed for. The clustering algorithm could detect the weaker signal at distances 
         not further than a few kpcs.} \label{Fig:ef1}
\end{figure}

\begin{figure}
\includegraphics[width=1.1 \linewidth]{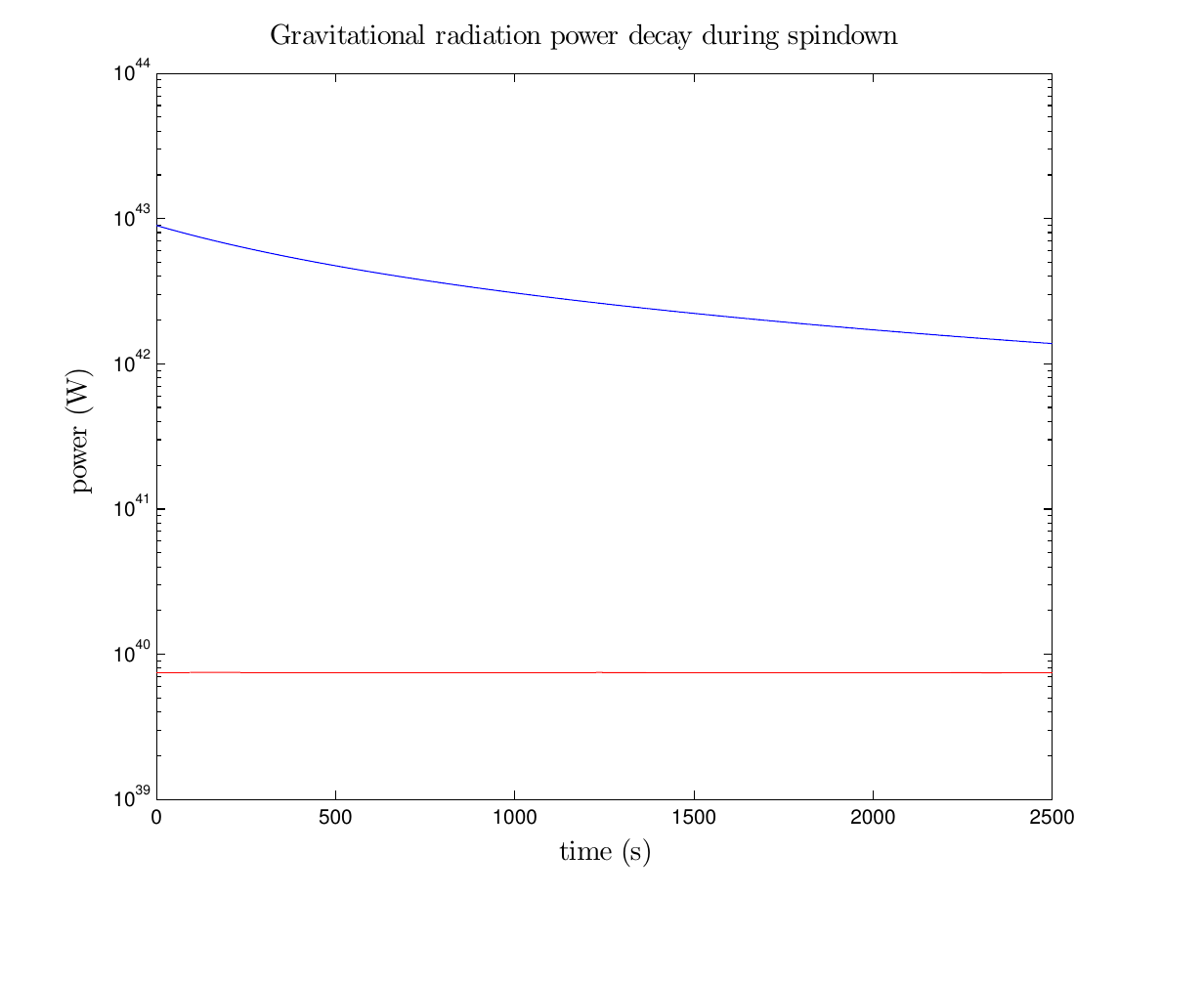}
\caption{These power evolutions correspond to the signals in Fig.\ref{Fig:ef1}. We see that the blue plot corresponds to a rapidly decaying signal: within $\unit[2500]{s}$ the 
radiation power drops to $17\%$ of its original value. The red plot corresponds to a power decay that dropped to only $99.9\%$ of the initial. The power of this (red) signal is about 
3 orders of magnitude lower than that of the signal plotted in blue. We chose this weak signal so that we can examine how the MLAs compare to the clustering algorithm both for powerful 
and weak signals.} \label{Fig:ef2}
\end{figure}

In the sensitivity study performed for the clustering algorithm, we used 9 distinct waveforms. These were chosen by taking ($\alpha$, $f_o$) pairs using
3 values ($10^{-1}$, $10^{-2}$, $10^{-3}$) for $\alpha$ and 3 values ($\unit[700]{Hz}$, $\unit[1100]{Hz}$, $\unit[1500]{Hz}$) for $f_o$. 
In this sensitivity study for the MLAs, for comparison purposes, we used 2 of these waveforms: $(f_o=\unit[1500]{Hz}, \alpha=0.1)$ and $(f_o=\unit[1100]{Hz}, \alpha=0.01)$. 
These waveforms as well as their corresponding power decays are shown in Fig.\ref{Fig:ef1} and Fig.\ref{Fig:ef2} respectively. 
MLAs are well suited especially for cases where the signal is not precisely (but only crudely) known. This paper is based on three specific MLAs: 
ANN \citep{haykin2004comprehensive}, SVM \citep{vapnik1998statistical} and CSC \citep{CSC_paper}. All three methods are considered novel applications 
in the area of long transient gravitational wave searches.  

This paper is organized as follows: In section \ref{Design} we present the details of the sensitivity study design. A  more detailed description about how the data is prepared is given 
in appendix \ref{Dataprep}. That discussion includes the resolution reduction performed on the data maps before we perform the training of the MLAs. After the training is performed we 
present the plot in Fig.\ref{Fig:ef11} of the training efficiencies as functions of the resolution reduction factor. This plot provides the motivation for reducing the resolution of the data maps 
by a factor of $10^{-2}$ per axis. In section \ref{MLAs} we present a summary of the three MLAs we used for our sensitivity study: Subsection \ref{ANNs} describes the training of the ANN 
algorithm, subsection \ref{SVMs} describes the training of the SVM and subsection \ref{CSCs} describes the training of the CSC algorithm. The details of the mathematical formulations are 
presented in the appendices: Appendix \ref{apANN} for the ANN, Appendix \ref{apSVM} for the SVM and Appendix \ref{apCSC} for the CSC. In section \ref{results} we present the results of our 
sensitivity study and compare the MLA efficiencies to the SC algorithm efficiencies. Finally, in section \ref{conclusions}, we summarize our conclusions and topics 
for future work.

\section{Sensitivity study design}
\label{Design}
  
The sensitivity study design for the MLAs has several differences from our previous sensitivity study for the clustering algorithms. For the latter we only had to produce 9 waveforms with 
$\alpha = 10^{-3}, 10^{-2}, 10^{-1}$ and $f_o = \unit[1500]{Hz}, \unit[1100]{Hz}, \unit[700]{Hz}$. For each waveform we created injection sets (100 injections per set) and each injection set 
corresponded to a specific injection distance. Using appropriate distance ranges we proceeded with this method until we got a $50\%$ detection rate. The distance corresponding to that 
success rate was marked as our detection distance. The detection threshold was taken to be the loudest `cluster SNR' \citep{STAMPPAPER} (as seen by the SC algorithm \citep{stochtrack}) 
among $1000$ frequency-time maps with pure detector noise. This threshold `cluster SNR' determined a false alarm probability (FAP) of 0.1\%.

The approach for the MLA sensitivity study was different. The MLAs were trained not only for those nine waveforms but also for as many as possible distinct waveforms. In the paragraphs that 
follow we discuss why we chose 11350 distinct waveforms each one injected at a distinct distance. Those 11350 waveforms had $\alpha$ and $f_o$ parameter values uniformly distributed in 
their corresponding parameter value ranges. By training the MLAs with these 11350 distinct injections we succeeded in getting the MLAs to recognize not only these 11350 waveforms but also all 
possible waveforms in the whole range of $\alpha$ and $f_o$ values at all possible distances (assuming the signal strength was high enough). This result of getting the MLAs recognizing signals 
outside the set of the signals used for training is called `generalization'.

\subsection{Choice of the $f_o$ and $\alpha$ parameter values}
                             
From equations \eqref{eq.1a} and \eqref{eq.2a} we have the two model parameters $\alpha$ and $f_o$ that determine the shape of the waveform. Apart from the shape, 
the injections that were produced for the training of the MLAs were also dependent on the pixel brightness or pixel signal-to-noise ratio (SNR). For a single pixel 
in the frequency-time maps (ft-maps) the SNR satisfies \citep{STAMPPAPER}
\begin{equation}
\label{eq.4_1_13}
 \text{SNR}(t,f, \hat{\Omega}) \propto Re \left [ \hat{Q}_{ij}(t,f, \hat{\Omega})  C_{ij}(t,f)  \right]   
\end{equation}

\noindent
where $i=1$ and $j=2$ are indices corresponding to the two advanced LIGO (aLIGO) detectors \citep{abbott2009ligo}, \citep{advancedLIGO} $\hat{Q}_{ij}(t,f, \hat{\Omega})$ 
is a filter function that depends on the source direction, $\hat{\Omega}$, \citep{abbott2005first} and $C_{ij} \equiv 2 \tilde{h}_i^*(t,f) \tilde{h}_j(t,f)$ is the cross 
spectral density, $\tilde{h}$ being the Fourier transform of the gravitational wave strain amplitude $h$. The latter is expressed in \citep{HOLAI} as a function of the 
distance $d$ to the source, the gravitational-wave frequency $f$ and the r-mode oscillation amplitude $\alpha$, by
\begin{equation}
\label{eq.3a}
 h \approx 1.5 \times 10^{-23}  \left( \frac{1\text{Mpc}}{d} \right)  \left( \frac{f}{1\text{kHz}}  \right )^3 | \alpha | . 
\end{equation}

\noindent
For the construction of the injection maps we chose the 3 parameter values $f_o$, $\alpha$ and $h^2$ to be uniformly distributed within predetermined value ranges as explained below.

Each injection set that was produced and used for the MLA training was limited to 11350 injection maps and 11350 noise maps. This was due to the 
computational resources available as well as the time needed to produce the 22700 maps. For each injection the waveform was randomly chosen in 
such a way that the $\alpha$ value was randomly chosen from a uniform distribution of 11350 $\alpha$ values in the range of $10^{-3}- 10^{-1}$, the 
$f_o$ value was randomly chosen from a uniform distribution of 11350 $f_o$ values in the range of $\unit[600-1600]{Hz}$, and for the $h^2$ values we 
picked 3 ranges (for 3 separate MLA trainings), whose choice is discussed in the next subsection, \ref{h_sq}.         
    
\subsection{Choice of the $h^2$ parameter values}
\label{h_sq}

The results of the sensitivity study for the clustering algorithm showed that for a signal of $f=\unit[1500]{Hz}$ and $\alpha=0.1$ the detection distance was 
up to $\unit[1.2]{Mpc}$. Using equation \eqref{eq.3a} we see that the SC algorithm can detect gravitational-wave strains of value $h \approx 4 \times 10^{-24}$. 
Values of the same order are obtained if we substitute the results for the other 8 waveforms. For example from table 1 in \citep{rmodespaper} we see that for 
$f=\unit[700]{Hz}$ and $\alpha=0.01$ we get a detection distance of $\unit[0.043]{Mpc}$. Substituting in equation \eqref{eq.3a} we get $h=1.2 \times 10^{-24}$. 
Therefore, the value of $h \approx 10^{-24}$ will become a reference point because this is the value of gravitational wave strain the MLAs will have to detect 
if they are shown to be at least as efficient as the SC algorithm \citep{stochtrack}.  

If we consider supernova events at distances in the range from $\unit[1]{kpc}$ to $\unit[1]{Mpc}$ then the corresponding range for the gravitational wave strain 
values is $h \approx 10^{-24}$ to $10^{-21}$. Therefore, there are several approaches in determining the range of $h^2$ values for the injection maps produced for 
the training of the MLAs. The first approach was to produce one set of data with injections at distances distributed in such a way that the $h^2$ values are uniformly 
distributed in the range: \\

\noindent
     (a) from $10^{-48}$ to $10^{-42}$ \,\,\,\,\ ($10^{-24} \le h \le 10^{-21} $ ). \\

In case the 11350 noise maps plus the 11350 injection maps will not be sufficient to achieve `generalization' during the training of the MLAs in the above range of 
values of $h$, the alternative steps would be to create injections with values of $h$ in smaller ranges. Therefore, we chose to produce three sets of data such that 
the $h^2$ values are uniformly distributed in the following ranges: \\

\noindent
     (b) from $10^{-46.4}$ to $10^{-45.4}$ \,\,\,\,\ ($10^{-23.2} \le h \le 10^{-22.7}$ ) \\
     (c) from $10^{-47.4}$ to $10^{-46.4}$ \,\,\,\,\ ($10^{-23.7} \le h \le 10^{-23.2}$ ) \\ 
     (d) from $10^{-48.0}$ to $10^{-47.4}$ \,\,\,\,\ ($10^{-24.0} \le h \le 10^{-23.7}$ ). \\
     
\noindent
The last choice of $10^{-24}$ is such that the waveform with $(f_o=\unit[1500]{Hz}, \alpha=0.1)$ may be detectable up to a distance of $\unit[5]{Mpc}$, 
depending on the MLA detection efficiencies. Note that at those distances (in the neighborhood of Milky Way) the supernova event rate is $1$ every 1-2 years \citep{galacticR, allrates}.  

After producing the simulated data using the stochastic pipeline, the noise and injection maps represent data in the frequency-time domain. Hence each data map is called an ft-map. 
These ft-maps data (already normalized) are preprocessed further so that we can create the `data matrix' that will include all the data that will be imported and used for the MLA 
training. Each row in the data matrix corresponds to an ft-map and each column corresponds to each pixel of the ft-map. The details of this preprocessing are explained in Appendix \ref{Dataprep}.
For the discussion on the MLAs that follows we will refer to our data used for the MLA training as the `data matrix'.

\section{Machine learning algorithms}
\label{MLAs}
There has only been a couple of applications of MLAs in the area of the detection of gravitational waves. One application is for gravitational wave searches of black hole binary coalescence 
with an application of random forests algorithms (RFA) \citep{bbcRF} and the other one is about the identification of noise artifacts (or glitches) in gravitational wave data with an application 
of ANN, SVM and RFA \citep{glitchesMLA}. Our study is investigating the application of three MLAs (ANN, SVM and CSC) for the detection of long transient gravitational wave signals. These are 
signals of duration from O(seconds) up to O(months) and they are in the middle of the spectrum (in terms of duration) between gravitational wave bursts and continuous waves.    
ANNs have been extensively studied and established \citep{basu2010use}. Similarly SVM \citep{cristianini2000introduction} and CSC \citep{CSC_paper} have been broadly applied. 

For the investigation of all three MLAs, the full set of data we had available was split into a $90\%$ for the training set and a $10\%$ for the test set. The training efficiencies 
mentioned in Fig.\ref{Fig:ef4}, Fig.\ref{Fig:ef5} and Fig.\ref{Fig:ef6} as well as in section \ref{results} are all referring to how well the MLAs can detect signals in the `unknown' 
$10\%$ of data that was left out of the training process. Before the selection of the training and test sets all data was shuffled and then the training and test sets were randomly 
selected.

\subsection{Artificial neural network}
\label{ANNs}

The applicability of an artificial neural network was investigated as a pattern recognition algorithm \citep{bishop1995neural} for the detection of r-mode gravitational waves. 
The aim was to train the ANN (Appendix \ref{apANN}) in order to make it capable of recognizing the ft-maps that contain r-mode signals and the ft-maps that contain pure detector 
noise. If successful for the r-mode gravitational-wave searches, the applicability of the ANNs may also be investigated in the stochastic pipeline for the detection of other 
long-lived gravitational-wave transients. 

The data matrix we used is a $2N \times d$ matrix where $N=11350$ is the number of data (ft-maps) with simulated detector noise. This is equal to the number ($N=11350$) of simulated data (ft-maps) with noise 
plus injected signals. According to Apppendix $\ref{Dataprep}$, the number of columns of the data matrix is chosen to be $d=550$ and this is equal to the dimensionality of the input layer $d=550$. The dimensionality 
of the hidden layer is $K=50$ and the dimensionality of the output layer is $L=2$. The `hidden' layer used `neurons' with the logistic sigmoid function \ref{eq.4a} while the output layer used neurons with the 
soft-max activation function \ref{eq.6a} which is typically used in classification problems to achieve a 1-to-n output encoding \citep{murphy2012machine, Bishop06a}. 

Using the above data matrix we performed a batch training of the neural network. After experimenting with various parameter populations we used a learning rate of $0.02$ and a momentum of $0.9$. For the training 
we used a batch version of the gradient descent as the optimization algorithm. To avoid over-fitting and maintain the ability of the network to `generalize' we used the `early stopping' technique. The results
as shown in Fig.\ref{Fig:ef4}, Fig.\ref{Fig:ef5} and Fig.\ref{Fig:ef6} demonstrate that the ANN algorithms performance is at least as good as that of the SC algorithm.

\subsection{Support vector machine}
\label{SVMs}

The second MLA we trained is a support vector machine (SVM). This method is based on a well formulated and mathematically sound theory \citep{Bishop06a}. The mathematical formulation of this algorithm is described in detail 
in Appendix \ref{apSVM}. In the SVM formulation we treat the noise ft-maps, as rows of a $N \times d$ matrix $X'_1$ and ft-maps with r-mode injections as rows of a $N \times d$ matrix $X'_2$ as points in a $d$-dimensional space. 
The solution to the SVM optimization problem is to find the optimal hypersurface that would separate (and hence classify) the noise points from the injection points. 

The above problem is a convex optimization problem and is formulated in Appendix \ref{apSVM}. It is solved using a state of the art sequential minimal optimization solver, LIBSVM. In our case we assumed that the classification 
problem is a non-linear one hence we introduced the radial basis (kernel) function (RBF). The constant $\gamma$ was taken to be the default (by LIBSVM) value and equal to $1/d$. For the other parameter $C$ was estimated to have
an optimal value in the range of $10^3$ to $10^5$.

\subsection{Constrained subspace classifier}
\label{CSCs}

The third algorithm we used is a constrained subspace classifier (CSC) as explained in Appendix \ref{apCSC}. The separation of the two classes is based on projecting the noise data points (represented by the $N \times d$ matrix $X'_1$) 
onto a $d_1$-dim subspace and similalry projecting the injection data points (represented by the $N \times d$ matrix $X'_2$) onto a $d_2$-dim subspace, where $d_1=d_2 < d$. Choosing the right trade-off between optimality and speed we 
picked dimensionalities $d_1=d_2=100$ for some cases (most powerful injections) and $d_1=d_2=200$ for some others (weakest injections). The constraint of the problem is the relative orientation of the two subspaces that is determined 
by the parameter $C$. This parameter was chosen after doing several runs and it was found to take values between $10^4$ and $10^5$.

\section{Results and discussion}
\label{results}

When using the SC algorithm in \citep{rmodespaper} the false alarm probability (FAP) is easily controlled by adjusting the SNR threshold above which 
an ft-map is considered to include an r-mode signal. This is not the case for the MLAs we used where the FAP is given after the training is performed as part of the training output. For 
this reason, to draw fair comparisons, we adjusted the FAP of the SC algorithm to match the output FAP of the MLAs. In Fig.\ref{Fig:ef4}, Fig.\ref{Fig:ef5} and Fig.\ref{Fig:ef6}, 
the results of the SC algorithm are compared with the results of the ANN, SVM and CSC for the same FAP. 

The first attempt to train all three MLAs was done with data produced with $h$ taking values over the range of $10^{-24} \le h \le 10^{-21} $. Using this range of values for $h$ the MLAs 
did not outperform the SC algorithm. This was probably due to the fact that the number 11350 of distinct signals used for the training was too small for the MLAs to achieve generalization, 
hence the training efficiencies are too low. To avoid this the next steps involved training of the same number of data over smaller ranges of values of $h$.      

In Fig.\ref{Fig:ef4} we present the detection efficiency results for the SC algorithm and the three MLAs on the $(\alpha=0.1, f_o=\unit[1500]{Hz})$ waveform. 
The MLAs were trained with data produced with $h$ taking values over the range of $10^{-23.7} \le h \le 10^{-23.2}$. This range of $h$ values is such that it includes the distance $d^{\prime}$
at which the SC algorithm has a $50\%$ detection efficiency (for this particular waveform).
This implies that the MLAs were trained with signals injected at distances a little shorter than $d^{\prime}$ up to distances 1.5-2 times longer than $d^{\prime}$. This particular choice resulted 
in an MLA performance that is at least as good as that of the SC algorithm.
The training of the MLAs on this training set resulted in false alarm probabilities of 4\%, 5\% and 10\% for the ANN, SVM and CSC respectively. 

At the 50\% false dismissal rate (FDR), the ANN shows an increase of $\sim$ 17\% in the detection distance, from $\sim$ \unit{1.45}Mpc (of the SC algorithm dash-dot 
blue line) to $\sim$ \unit{1.80}Mpc (of the solid blue line). The SVM shows an increase of $\sim$ 23\%, from $\sim$ \unit{1.50}Mpc (of the SC algorithm dash-dot green 
line) to $\sim$ \unit{1.85}Mpc of the solid-green line. The CSC shows an increase of $\sim$ 13\%, from $\sim$ \unit{1.50}Mpc (of the SC algorithm dash-dot red line) 
to $\sim$ \unit{1.70}Mpc of the solid-red line.

In  Fig.\ref{Fig:ef5} we present the detection efficiency results for the SC algorithm and the three MLAs on the $(\alpha=0.1, f_o=\unit[1500]{Hz})$ waveform. 
The latter were trained with data produced with $h$ taking values over the range of $ 10^{-24.0} \le h \le 10^{-23.7}$. 
This range of $h$ values is such that all of the injection distances of the training set were higher than the distance $d^{\prime}$, at which the SC algorithm has a $50\%$ detection 
efficiency (for this particular waveform). The training of the MLAs with injections at distances longer than $d^{\prime}$ was done in order to push the limits of the MLAs and see how much (if at all) 
they can outperform the SC algorithm. 

The training of the MLAs on this training set resulted in high false alarm probabilities of 18\%, 22\% and 36\% for the ANN, SVM and CSC respectively. 
At the 50\% FDR, the ANN algorithm shows an increase of $\sim$ 137\% in the detection distance, from \unit{1.50}Mpc (of the SC algorithm dash-dot blue) to \unit{3.55}Mpc 
(of the solid blue). The SVM algorithm shows an increase of $\sim$ 83\% in the detection distance, from \unit{1.50}Mpc (of the SC algorithm dash-dot green) to \unit{2.75}Mpc 
(of the solid green). The CSC shows an increase of $\sim$ 46\% in the detection distance, from $\sim$ \unit{2.60}Mpc (of the SC algorithm dash-dot red line) to 
$\sim$ \unit{3.40}Mpc (of the solid red line). The distance range covered in this set has a practical significance because it covers: (a) the distance of $\unit[3.5]{Mpc}$ 
at which the January 2014 supernova occured in M82 and (b) the distance of $\unit[5]{Mpc}$ at which the supernova event rate in the Milky Way neighborhood is about 1 every 1-2 years. 

In Fig.\ref{Fig:ef6} we present the detection efficiency results for the SC algorithm and the three MLAs on the $(\alpha=0.01, f_o=\unit[1100]{Hz})$ waveform. The MLAs were trained 
with data produced with $h$ taking values over the range of $10^{-24} \le h \le 10^{-23.7}$. This range of $h$ values is such that it includes the distance $d^{\prime}$
at which the SC algorithm has a $50\%$ detection efficiency (for this particular waveform).
This implies that the MLAs were trained with signals injected at distances a little shorter than $d^{\prime}$ up to distances 1.5-2 times longer than $d^{\prime}$. This particular choice resulted 
in an MLA performance that is not as good as our results for the $(\alpha=0.1, f_o=\unit[1500]{Hz})$ waveform.
The training of the MLAs on this training set resulted in false alarm probabilities of 18\%, 22\% and 36\% 
for the ANN, SVM and CSC respectively. At the 50\% false dismissal rate (FDR), the ANN shows an increase of $\sim$ 18\% in the detection distance, from $\sim$ \unit{170}kpc (of the SC algorithm dash-dot blue line) 
to $\sim$ \unit{210}kpc (of the solid blue line). The SVM shows a decrease of $\sim$ 24\%, from $\sim$ \unit{170}kpc (of the SC algorithm dash-dot green line) to $\sim$ \unit{140}kpc 
of the solid-green line. The CSC shows a decrease of $\sim$ 3\%, from $\sim$ \unit{185}kpc (of the SC algorithm dash-dot red line) to $\sim$ \unit{10}kpc of the solid-red line.

\begin{widetext}

\begin{sidewaysfigure}[htbp!]
\centering
\includegraphics[width=0.95 \linewidth]{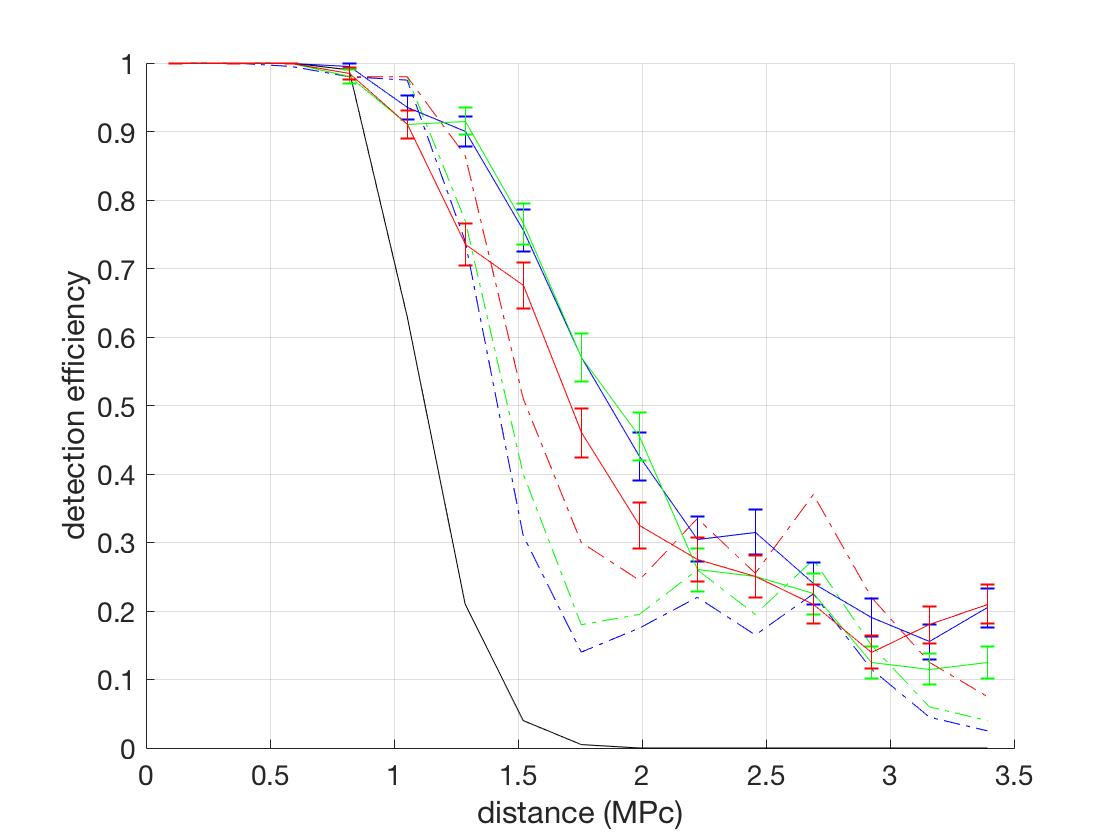} 
% {table_III_av-eps-converted-to.pdf}
\caption{These are the detection efficiencies for the $(f_o=\unit[1500]{Hz}, \alpha=0.1)$ waveform binned every $0.2$ kpc. This waveform was the most powerful signal we used in our previous sensitivity study \citep{rmodespaper}. 
         The dash-dot black line corresponds to the SC algorithm sensitivity study at FAP equal to 0.1\%.
         The (blue, green and red) dash-dot lines correspond to the detection efficiencies of the SC algorithm at FAP of 4\%, 5\% and 10\%. The (blue, green and red) 
         solid lines correspond to the detection efficiencies of the ANN (at 4\%), SVM (at 5\%) and CSC (at 10\%) respectively. This plot demonstrates that (when compared for 
         the same FAP) the MLAs performance is at least as good as that of the SC algorithm. At the 50\% false dismissal rate (FDR), the ANN shows an increase of 
         $\sim$ 17\% in the detection distance - from $\sim$ \unit{1.45}Mpc (of the SC algorithm dash-dot blue line) to $\sim$ \unit{1.80}Mpc (of the solid blue line). 
         The SVM shows an increase of $\sim$ 23\% - from $\sim$ \unit{1.50}Mpc (of the SC algorithm dash-dot green line) to $\sim$ \unit{1.85}Mpc of the solid-green line. 
         The CSC shows an increase of $\sim$ 13\% - from $\sim$ \unit{1.50}Mpc (of the SC algorithm dash-dot red line) to $\sim$ \unit{1.70}Mpc of the solid-red line.}\label{Fig:ef4}
\end{sidewaysfigure}

 \begin{sidewaysfigure}[htbp!]
\centering
\includegraphics[width=0.95 \linewidth]{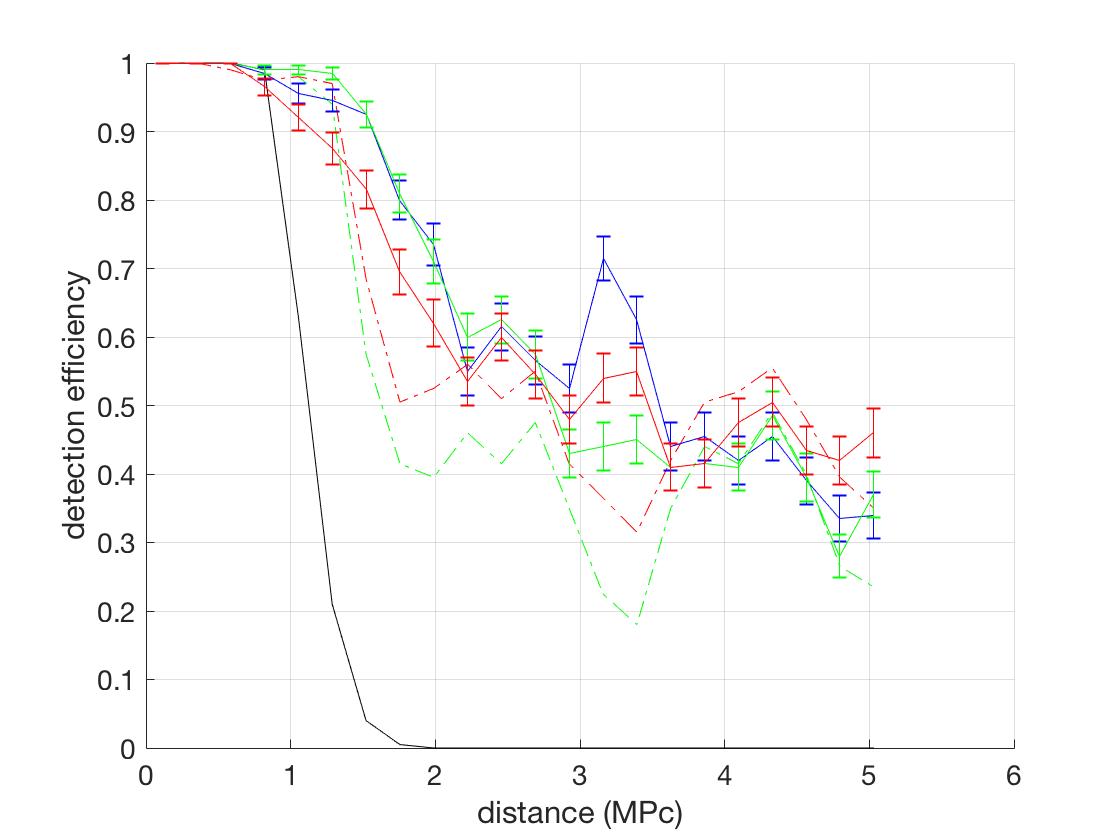}
% {table_IV_av-eps-converted-to.pdf}
\caption{These are the detection efficiencies for the $(f_o=\unit[1500]{Hz}, \alpha=0.1)$ waveform binned every $0.2$ kpc. This waveform was the most powerful signal we used in our previous sensitivity study \citep{rmodespaper}.
         The dash-dot black line corresponds to the SC algorithm sensitivity study at FAP equal to 0.1\%. 
         The (blue, green and red) dash-dot lines correspond to the detection efficiencies of the SC algorithm at FAP of 18\%, 22\% and 36\% respectively,  though the detection efficiencies at 18\% and 22\% do not differ, so the lines overlap. The (blue, green and red) 
         solid lines correspond to the detection efficiencies of the ANN (at 18\%), SVM (at 22\%) and CSC (at 36\%) respectively.
         At the 50\% FDR, the ANN algorithm shows an increase of $\sim$ 137\% in the detection distance - from \unit{1.50}Mpc (of the 
         SC algorithm dash-dot blue) to \unit{3.55}Mpc (of the solid blue). The SVM algorithm shows an increase of $\sim$ 83\% in the detection distance - from \unit{1.50}Mpc 
         (of the SC algorithm dash-dot green) to \unit{2.75}Mpc (of the solid green). The CSC shows an increase of $\sim$ 46\% in the detection distance 
         - from $\sim$ \unit{2.60}Mpc (of the SC algorithm dash-dot red line) to $\sim$ \unit{3.40}Mpc (of the solid red line). 
         The distance range covered in this set has a practical significance because it covers: (a) the distance of $\unit[3.5]{Mpc}$ at which the January 
         2014 supernova occured in M82 and (b) the distance of $\unit[5]{Mpc}$ at which the supernova event rate in the Milky Way neighborhood is about 1 every 1-2 years.}\label{Fig:ef5}
\end{sidewaysfigure}

\begin{sidewaysfigure}[htbp!]
\centering
\includegraphics[width=0.95 \linewidth]{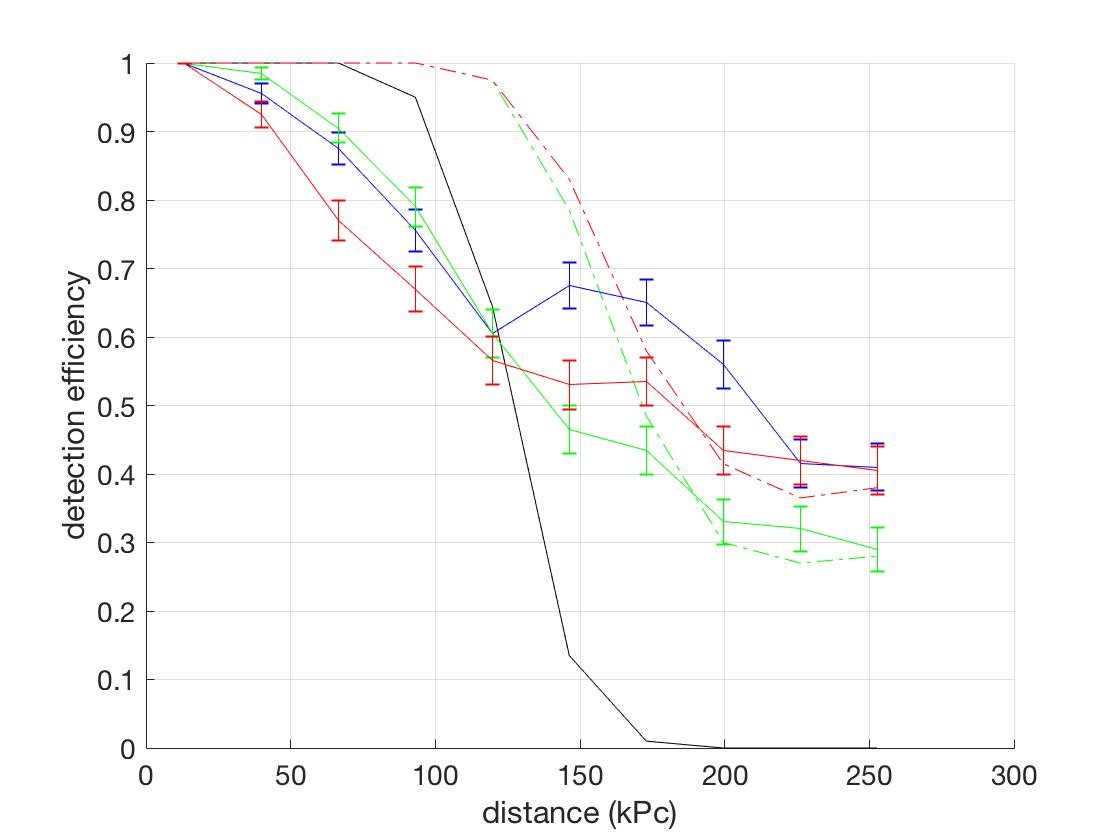}
% {table_VI_av-eps-converted-to.pdf}
\caption{These are the detection efficiencies for the $(f_o=\unit[1100]{Hz}, \alpha=0.01)$ waveform binned every $0.2$ kpc. This signal was proven (in our previous study) to be detectable only at distances 
         that cover the Milky Way. This signal is approximately monochromatic for the durations our sensitivity studies were designed. 
         The (blue, green and red) dash-dot lines correspond to the detection efficiencies of the SC algorithm at FAP of 18\%, 22\% and 36\% respectively, though the detection efficiencies at 18\% and 22\% do not differ, so the lines overlap. The (blue, green and red) 
         solid lines correspond to the detection efficiencies of the ANN (at 18\%), SVM (at 22\%) and CSC (at 36\%) respectively.
         At the 50\% false dismissal rate (FDR), the ANN shows an increase of $\sim$ 18\% in the detection distance - from $\sim$ \unit{170}kpc (of the SC algorithm dash-dot blue line) 
         to $\sim$ \unit{210}kpc (of the solid blue line). The SVM shows a decrease of $\sim$ 24\% - from $\sim$ \unit{170}kpc (of the SC algorithm dash-dot green line) to $\sim$ \unit{140}kpc 
         of the solid-green line. The CSC shows a decrease of $\sim$ 3\% - from $\sim$ \unit{185}kpc (of the SC algorithm dash-dot red line) to $\sim$ \unit{10}kpc of the solid-red line. }\label{Fig:ef6}
\end{sidewaysfigure}

\end{widetext}

\begin{figure}[htbp!]
  \centering
  \begin{tabular}{@{}p{1.0\linewidth}@{\quad}p{1.0\linewidth}@{}}
    \subfigimg[width=1.0 \linewidth]{}{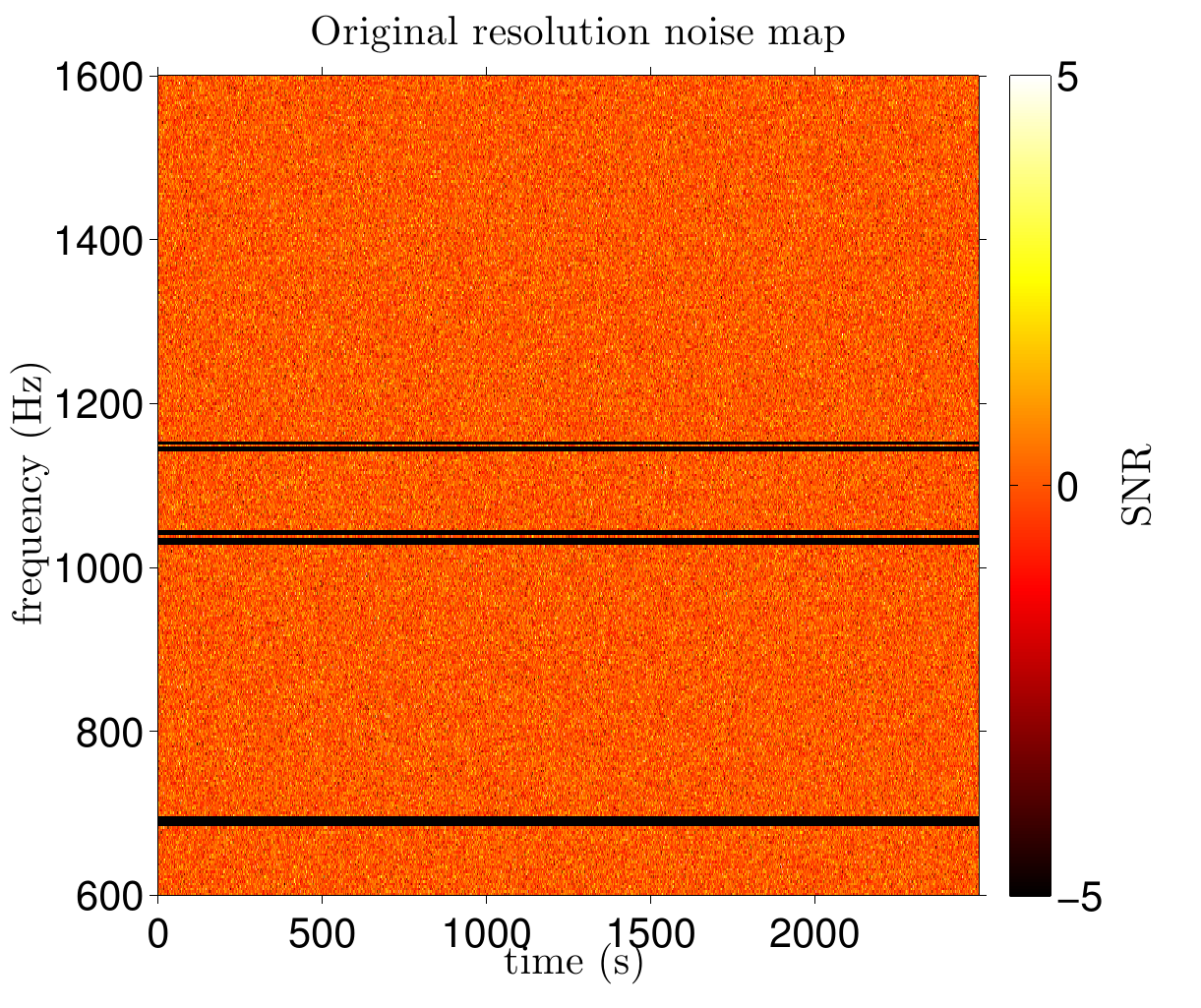} \caption{This is one of the noise ft-maps with the original resolution of $1000 \times 5000$ pixels. The 
                                                                pixels along the vertical axis correspond to $1$Hz each. The pixels along the horizontal axis 
                                                                correspond to $0.5$s each, hence the total duration of the map is 2500s. The frequency cuts are well known
                                                                seismic frequency bands and suspension vibration modes.
                                                                }\label{Fig:ef7} &                                                   
    \subfigimg[width=1.0 \linewidth]{}{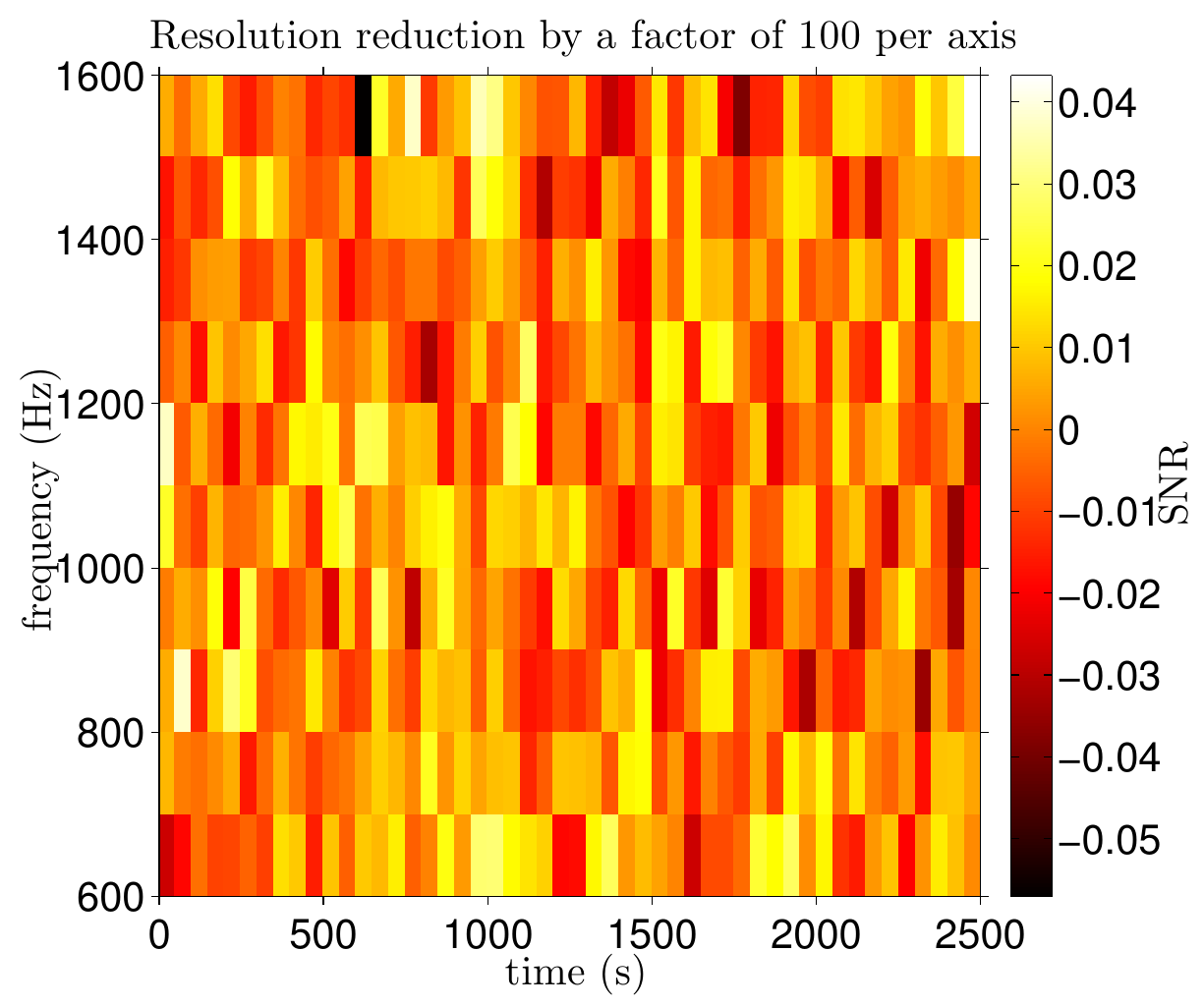} \caption{The highest training efficiency for the MLAs was achieved with resolution reduction by a factor of 100 per axis, 
                                                                   as seen in Fig.\ref{Fig:ef11}. This reduced $10 \times 50$ resolution ft-map corresponds to the full resolution noise map in Fig.\ref{Fig:ef7}. 
                                                                   For the resolution reduction we used bicubic interpolation as provided by the matlab imresize.m function. The 
                                                                   frequency cuts were substituted with zeros before reducing the resolution.
                                                                   }\label{Fig:ef8} \\
    \subfigimg[width=1.0 \linewidth]{}{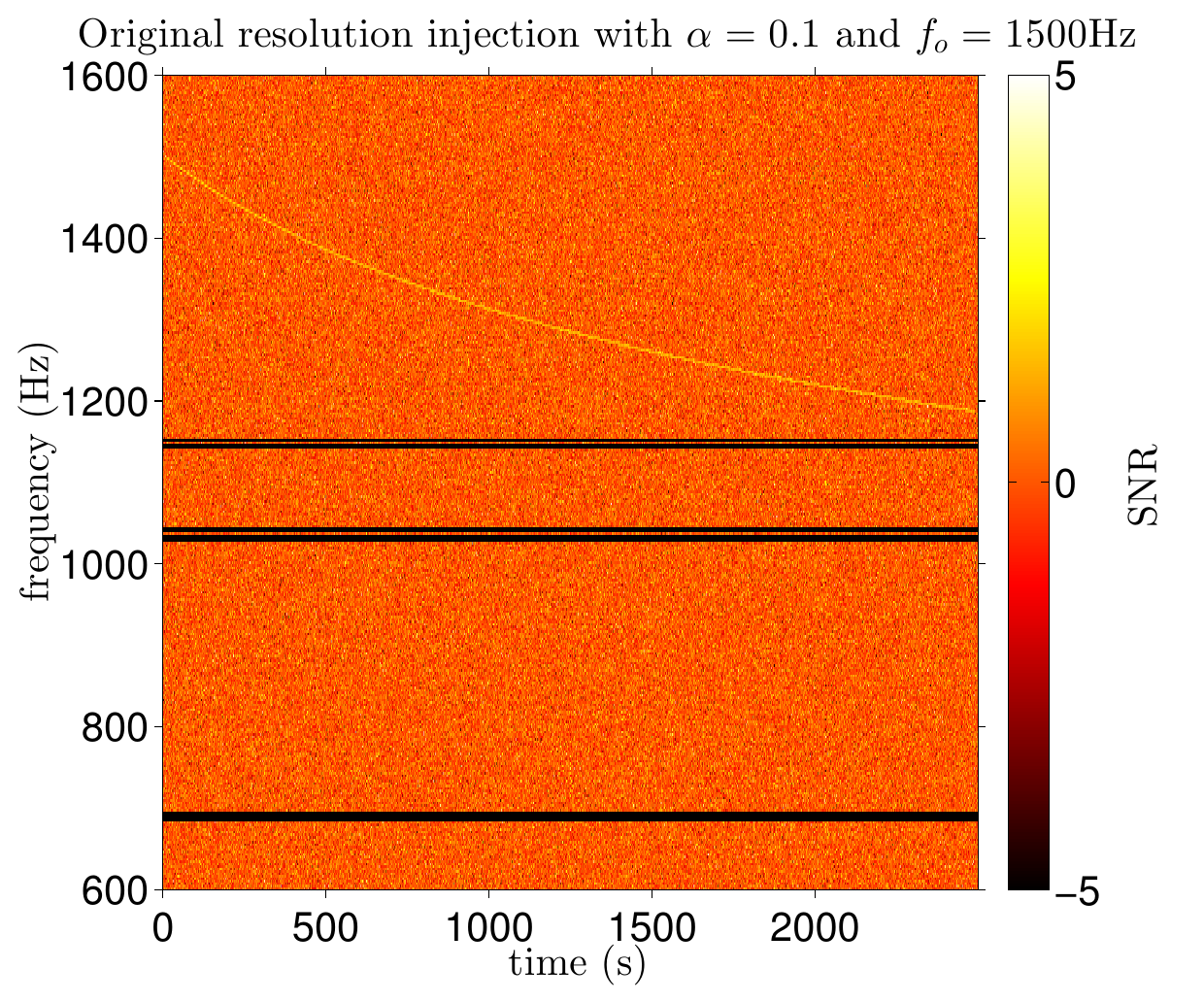}\caption{This is an injection added to the noise ft-map shown in Fig.\ref{Fig:ef7}. The waveform has parameters $\alpha=0.1$ and $f_0=1500$ Hz.  
                                                             The duration of the injection is $2500$s and corresponds to a distance to the source of $\unit[117]{kpc}$. Injections at longer
                                                             distances are harder to see by eye in the original resolution maps. The contrast between signal pixels and noise pixels
                                                             is higher in the reduced resolution maps as shown in Fig.\ref{Fig:ef8}. This makes it easier to see the injections in the reduced 
                                                             resolution maps rather than the full resolution ft-maps.
                                                             }\label{Fig:ef9} &
    \subfigimg[width=1.0 \linewidth]{}{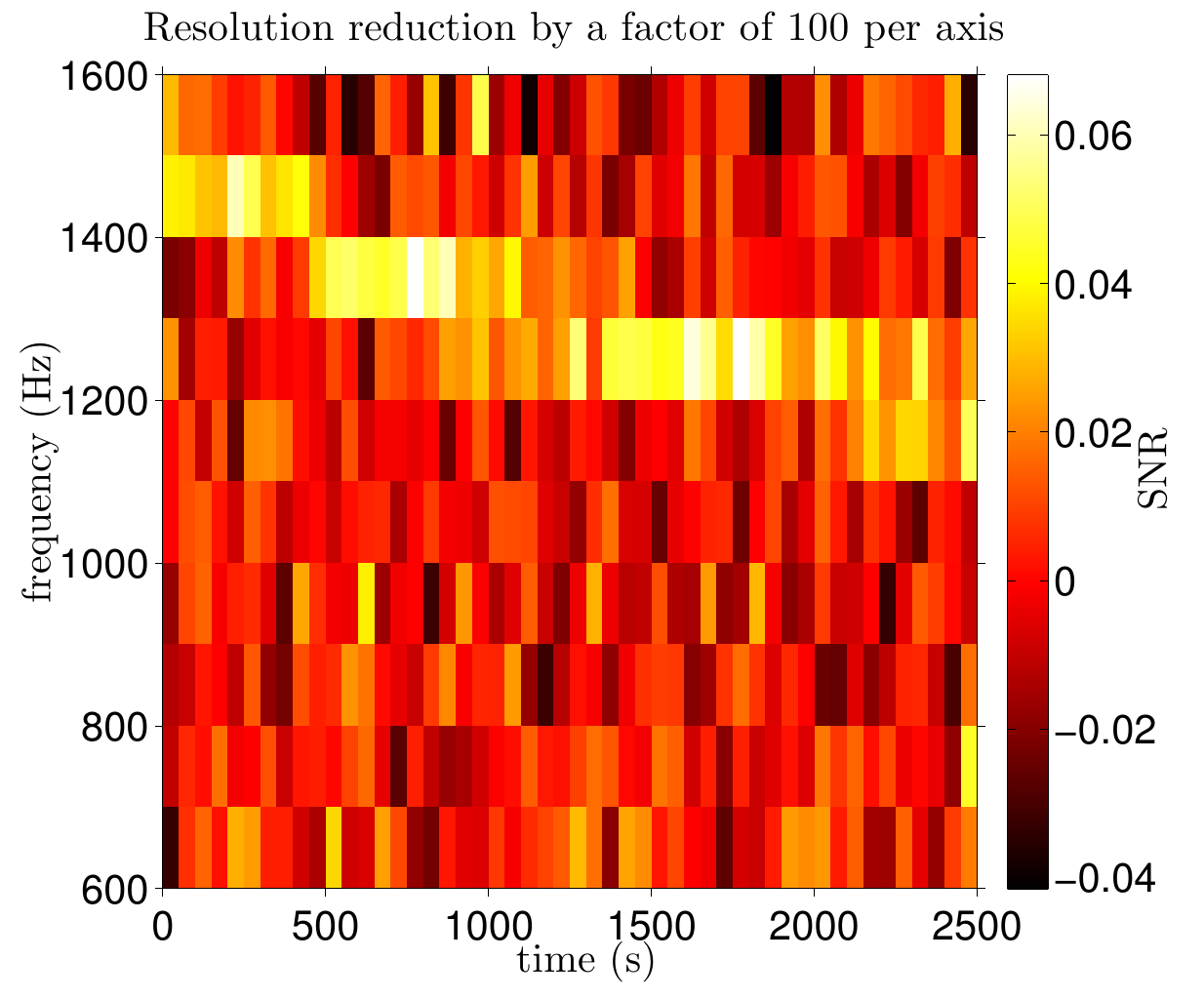}\caption{This reduced $10 \times 50$ resolution ft-map corresponds to the full resolution map in Fig.\ref{Fig:ef9}. Despite the 10000 times 
                                                             reduced resolution as compared to the ft-map of Fig.\ref{Fig:ef9}, the r-mode injection is still visible. It turns out that the reduced 
                                                             resolution ft-maps increase the training efficiency for the MLAs, according to Fig.\ref{Fig:ef11}. However, for the parameter estimation 
                                                             algorithms we use the full resolution ft-maps.  
                                                             }\label{Fig:ef10}
  \end{tabular}
  
\end{figure}

\section{Conclusions}
\label{conclusions}

Computational efficiency: The most computationally expensive part of this study was the production of the one set of 11350 noise ft-maps and the 3 sets of 11350 
injection ft-maps (each set requires up to 10 GB of memory and up to 1 week on a 50 node cluster). The 3 sets of injections examined the 3 different ranges of values 
for $h$ (those correspond to 3 different ranges of values for the distance). In practice, we will know the distance to the source so we will have to produce only one 
set of injections that will be determined according to that distance. 
 
Training/testing speeds: Once we have the method (that is presented in this paper) the training of the CSC method requires 10 minutes, the training of the SVM 
method requires about 30 minutes while the training of the ANN method requires about 8 hours. After the training is done the decision making about the presence of a
signal or not takes about 2 seconds for 100 ft-maps. The MLAs are much faster when it comes to the decision making process than the SC algorithm is (that takes
up to 5 minutes for one ft-map).
 
Detection performance: Fig.\ref{Fig:ef4} shows the detection efficiencies of MLAs that were trained with signals injected at distances a little shorter than the distance 
$d^{\prime}$, at which the SC algorithm has a $50\%$ success rate, up to distances 1.5-2 times longer than $d^{\prime}$. When compared to Fig.\ref{Fig:ef5} that 
shows the detection efficiencies of MLAs trained with signals injected at distances longer than $d^{\prime}$ (from 2.2 up to 4.3 times longer) we observe that the MLAs of Fig.\ref{Fig:ef4}
do not perform as well. In both figures the MLAs outperform the SC algorithm by a factor of 1.2 (Fig.\ref{Fig:ef4}) up to a factor of 1.8 (Fig.\ref{Fig:ef5}).  
Training the MLAs with injections at distances shorter than $d^{\prime}$ was to ensure that the MLAs can detect signals injected at distances $0.7-0.8$ that of $d^{\prime}$, and 
training the MLAs with injections at distances longer than $d^{\prime}$ was done in order to push the limits of the MLAs and see how much (if at all) they can outperform the SC algorithm.                      
 
Low detection efficiency: for the $(0.01, \unit[1100]{Hz})$ waveform. We suspected that the low detection efficiencies for the second waveform (weakest signal) as seen in Fig.\ref{Fig:ef6} 
are due to the resolution reduction factor of $10^{-2}$ we used. This resolution reduction factor was shown (in Fig.\ref{Fig:ef11}) to maximize the training efficiency for the strongest signals
(Fig.\ref{Fig:ef4} and Fig.\ref{Fig:ef5}) $(0.1,\unit[1500]{Hz})$. We did not derive the optimal value of this reduction factor for the weaker signals. In other words, we have not tested whether 
the weaker signals have maximum training efficiencies at a different resolution reduction than the one we used for the strongest signal. This needs further investigation. 
 
False alarm probabilities: In our study FAPs of 4-10$\%$ (for Fig.\ref{Fig:ef4}) and 18-36\% (for Fig.\ref{Fig:ef5} and Fig.\ref{Fig:ef6}) are considered very high, however, a more carefully chosen 
training set may result in lower FAPs. The first suggestion would be to train the MLAs with a higher number of noise and injection ft-maps. If that is not possible (due to data availability) 
we may train the MLAs with injections at distances over a range of ($h^2$) values that is smaller than those in the current training sets. Similarly we can use smaller ranges of 
parameter values for $\alpha$ and $f_o$. We can also try to increase the ratio of noise maps over injection maps in the training set so that the MLAs may recognize the noise maps 
more efficiently. Specifically for the ANN, one way we may try to reduce the FAP is by exploring different topologies in the neural network architecture. For SVM and CSC we may 
introduce a cost function to suppress FAP to acceptable values. 

For the most powerful signals the false alarm probability is about 3\%. This FAP is further decreased down to 0.3\% when we used a number of noise maps 2 times higher 
than the number of signal-injection maps. However, this was done at the expense of the True Positive probability (that decreased from 99\% to 96\%). Therefore, for 
signals from nearby sources the MLAs were shown to have a FAP comparable to what the referee would like to see.  

Search optimization: There are many ways that we can further optimize the MLAs specifically designed for the search of r-mode gravitational radiation. One way is by customizing 
the ft-map resolution reduction. Instead of using bicubic interpolation we may use a resolution reduction algorithm specifically designed for the r-mode signals so that the averaging 
is done along the r-mode signal curves. Since the r-mode search is a targeted search (using a supernova electromagnetic or neutrino trigger) the distance to the source can be estimated 
with an accuracy of $10-15\%$ \citep{TYPE1DIST,TYPEIIDIST}. This distance range can then be used to produce injection ft-maps with which the MLAs will be trained. In this way the training
can be optimized for the distance of the detectors to the external trigger. 
 
Search constraints: Our current method is specifically designed for r-mode gravitational wave searches. A different signal (e.g. gravitational waves sourcing from other neutron 
star oscillation modes) would require their own training set produced over the specific model parameter values. This is a quite different approach than that of the 
SC algorithm that is generically designed for the detection of any type of signal. Our current method involves the production of at least 10000 ft-maps (that may be overlapping), any 
amount of data that will not be enough for the production of this many ft-maps will limit the sensitivity of the search. At the same time the higher the number of the ft-maps used 
for training is the more we may increase the training efficiencies of the MLAs.     
  
Resolution reduction: We did not examine robustness of the resolution reduction results on other signals (with different $\alpha$ and $f_o$) therefore, this method may have to be repeated for different r-mode waveforms as well as different long duration gravitational wave transients. However, if we use Google's ``Tensorflow'' \cite{tensorflow}, based on graphic processing units (GPUs) we may be able to train computationally expensive algorithms, such as region convolutional neural networks (R-CNNs), without needing to reduce the resolution of the time/frequency map.     

Despite the high FAP, the MLAs have an extremely important advantage over SC algorithm. After performing the training stage of the MLAs, the testing stage 
is lightning fast (testing with the SC algorithm may take several tens of minutes versus fractions of a second that are needed by MLAs). This implies that we can 
use the MLAs as part of an investigative stage in the pipeline that would be able to provide very fast and solid triggers for further, and more intense, investigation.

Pipeline suitability: ANN, SVM and CSC (and very likely other machine learning algorithms not tested yet) are a suitable class of decision making algorithms in 
the search not only for r-mode gravitational waves but in the search for long transient gravitational waves in general. The results in this paper demonstrate that
the stochastic pipeline would benefit from utilizing machine learning algorithms for determining the presence of a signal or not. 

The aim of this paper was not to demonstrate how we can use MLAs in order to make a detection announcement. Instead, our aim was to perform a preliminary investigation 
on how we can use raw data taken by the LIGO detectors, pre-process it and feed it in three separate MLAs. The purpose of this paper was fulfilled since we were able to 
obtain preliminary results that can encourage us (as well as other groups) for further investigation, including addressing the issue of high false alarm probability. 
 
\section{Suggestions for future work}
\label{future_work}
 
Future developments: Future developments include optimization of the current methods as well as the use of other supervised machine learning algorithms such as random 
forests\citep{breiman2001random}. Random forests can deal with the high dimensionality of our data by revealing features that contribute very low information to our analysis; 
which can be discarded prior to classification. With respect to the ANN, we plan to train a deep convolutional neural network \cite{krizhevsky2012imagenet} which appears 
to be very promising for image classification.                

Training with more data: Out of the 5 million columns (of our 22700 x 5000000 matrix) only the 22700 are linearly independent (row rank=column rank). This means that the 
information we can extract from the columns are limited by the number of data rows we produce. This suggests that upon production of higher number of data rows (ft-maps) 
the MLAs will be able to extract more information from the data matrix and quite possibly the training efficiencies will improve.  
 
\section*{Acknowledgments}

This work has been supported by grants PHY 0855313 and PHY 1205512 from the NSF. 
 
\appendix

\section{Artificial Neural Networks}
\label{apANN}

After resolution reduction the original $2N \times D$ data matrix gets reduced to a $2N \times d$ data matrix, $X'$. The latter will be presented as input into a 
feed-forward neural network with an input layer of dimensionality $d$. For the training of the ANN we randomly picked $90\%$ of the first $N$ (injection data) rows 
and also $90\%$ of the second $N$ (noise data) rows. The other $10\%$ of the (injection and noise) rows was used to determine the training efficiency of the trained 
algorithm. The ANN had one hidden layer with a number of nodes (`neurons') equal to $K$ and an output layer with two `neurons' that would `fire' for `signal' or 
`no signal'. The `hidden' layer used `neurons' with the logistic sigmoid function \citep{murphy2012machine}
\begin{equation}
\label{eq.4a}
 \sigma(a_j) =\frac{1}{1+\exp(-a_j)}
\end{equation}
 
\noindent
where $a_j$ ($ j= 1,2,...,d$) are the values presented at one `neuron' in the hidden layer. The purpose of the hidden layer is to allow for non-linear combinations of 
the input values to be forwarded to the output layer. These combinations in the hidden layer carry forward `features' from the input to the output layer 
that would not be possible to be extracted from each individual neuron in the input layer, enabling non-linear classification. The number of hidden layers 
and hidden neurons was chosen, as is typically done, after experimentation with various ANN architectures, aiming to enhance the accuracy, the robustness 
and the generalization ability of the ANN, along with the training efficiency and feasibility.

Starting from the first ft-map in the data matrix $X$ i.e. starting from the row vector $x_1$ where
\begin{equation}
\label{eq.x1}
 x_1= \{ \ x_{ij} | \ i=1 \,\ \mbox{and} \,\ j=1,2,...,d \}
\end{equation}

\noindent
we have $d$ values that are fed into the input layer of the neural network.
These values are then non-linearly combined in each hidden `neuron' to get $K$ many output values forwarded to the output layer, given by
\begin{equation}
\label{eq.5a}
 x'_{1k} = \sigma \large ( \sum^{d}_{j=1} w_{kj}^{(1)}x_{1j} + w_{k0}^{(1)} \large )
\end{equation}

\noindent
where $k=1,2,...,K$ is the index corresponding to each `neuron' in the hidden layer and the superscript (1) represents the hidden layer. The parameters 
$w_{kj}$ are called the weights while the parameters $w_{k0}$ are called the biases of the neural network. 

The `output' layer used `neurons' with the soft-max activation function which is typically used in classification problems to achieve a 1-to-n output encoding \citep{Bishop06a}. 
In particular, the soft-max function rescales the outputs in order for all of them to lie within the range $[0,1]$ and to sum-up to 1. This is done by 
normalizing the exponential of the input $b_k$ to each output neuron over the exponential of the inputs of all neurons in the output layer:
\begin{equation}
\label{eq.4softmax}
 \text{soft-max}(b_k) =\frac{\exp(b_k)}{\sum_k(\exp(b_k))} .
\end{equation} 

\noindent
When the values from equation \eqref{eq.5a} are presented in the output layer we get the result 
\begin{equation}
\label{eq.6a}
x''_{1l} = \mbox{soft-max} \large ( \sum^{K}_{k=1} w_{lk}^{(2)} x'_{1k} + w_{l0}^{(2)} \large )
\end{equation}
 
\noindent
(where $l=1,2$) as the output value in the single neuron of the output layer. Equation \eqref{eq.6a} represents the `forward propagation' of information in 
the neural network since the inputs are `propagated forward' to produce the outputs of the ANN, according to the particular `weights' and `biases'.

Equation \eqref{eq.6a} also shows that a neural network is a non-linear function, $\mathcal{F}$, from a set of input variables $\{x_{i}\}$ such that 
$i \in \{1,2,...,2N\}$ as defined by equation \eqref{eq.x1} i.e. $x_i$ are row vectors of the matrix $X'$ to a set of output variables $\{x''_l \}$ such that 
$l \in \{1,2 \}$ i.e. the output layer has dimensionality equal to $L=2$ (2 neurons: one fires for noise and the other fires for injection). To merge the weights $w^{(1)}_{kj}$ and 
biases $w^{(2)}_{k0}$ into a single matrix (and similarly do with the weights $w^{(2)}_{lk}$ and biases $w^{(2)}_{l0}$)  we need to redefine $x_1$ as 
given by equation \eqref{eq.x1} to  
\begin{equation}
\label{x1}
 x_1= \{ \ x_{ij} | \ i=1 \,\ \mbox{and} \,\ j=0,1,2,...,d \,\ \mbox{and} \,\ x_{10}=1 \}
\end{equation}

\noindent
and similarly redefine all row vectors of $X'$ as well as all the output row vectors from the hidden layer. Then the non-linear function $\mathcal{F}$ is controlled 
by a $(K+1) \times (d+1)$ matrix $\bold{w^{(1)}}$ and a $2 \times(K+1)$ matrix $\bold{w^{(2)}}$ of adjustable parameters. Training a neural network corresponds 
to calculating these parameters.   
   
Numerous algorithms for training ANN exist \citep{murphy2012machine} and in general can be classified as being either sequential or batch training methods: \\              
(i) sequential (or `online') training: A `training item' consists of a single row (one ft-map) of the data matrix. In each iteration a single row is passed through 
the network. The weight and bias values are adjusted for every `training item'  based on the difference between computed outputs and the training data target outputs. \\
(ii) batch training: A `training item' consists of the matrix $X'$ (all $2N$ rows of the data matrix). In each iteration all rows of $X'$ are successively passed through 
the network. The weight and bias values are adjusted only after all rows of $X'$ have passed through the network. 
  
In general, batch methods perform a more accurate estimate of the error (i.e. the difference between the outputs and the training data target outputs) and hence (with 
sufficiently small learning rate \citep{wilson2003general}) they lead to a faster convergence. As such, we used a batch version of gradient descent as the optimization algorithm. This 
form of algorithm is also known as `back-propagation' because the calculation of the first (or hidden) layer errors is done by passing the layer 2 (or output) layer 
errors back through the $w^{(2)}$ matrix. The `back-propagation' gradient descent for ANNs in batch training is summarized as follows:
\begin{algorithm} [H]                   
\caption{Gradient Descent for ANN}   
\label{gradient_descent}                           
\begin{algorithmic}               
\begin{small}
\STATE 1. Initialize $w$ (and biases) randomly. 
\WHILE{error on the validation set satisfies certain criteria} 
   \FOR{i=1:2N}
   \STATE 2. Feed-forward computation of the input vector $x_i$.
   \STATE 3. Calculate the error at the output layer.
   \STATE 4. Calculate the error at hidden layer.     
   \STATE 5. Calculate the mean error.  
   \STATE 6. Update $w$ of the output layer.
   \STATE 7. Update $w$ of the hidden layer.
   \ENDFOR
\ENDWHILE
   \end{small}
\end{algorithmic}
\end{algorithm}

Out of the $90\%$ of the data that was (randomly) chosen for the training, $10\%$ of that was used as a validation set. The latter is used in the 
`early stopping' technique that is used to avoid over-fitting and maintain the ability of the network to `generalize'. Generalization is the ability 
of a trained ANN to identify not only the points that were used for the training but also points in between the points of the training set. For each 
iteration the detection efficiency of the ANN is tested on the validation set. When the error on the validation set drops by less than $10^{-3}$ for 
two consecutive iterations then we do the `early stopping' and the training is stopped.  

The learning rate of the gradient-decent algorithm determines the rate at which the training of the network is moving towards the optimal parameters. 
It should be small enough not to skip the optimal solution but large enough so that the convergence is not too slow. A crucial challenge for the algorithm 
is not to converge to local minima. This can be avoided by adding a fraction of a weight update to the next one. This method is called `momentum' of the 
training of the network. Adding `momentum' to the training implies that for a gradient of constant direction the size of the optimization steps will increase. 
As such, the momentum should be used with relatively small learning rate in order not to skip the optimal solution.

\section{Support Vector Machine}
\label{apSVM}

The second MLA we trained is a support vector machine (SVM). This method gained popularity over the ANNs because it is based on well formulated and 
mathematically sound theory \citep{Bishop06a}. In the following paragraphs we give a brief introduction to the SVM mathematical formulation. 

In the SVM formulation we treat the noise ft-maps, rows of $X'_1$ as well as the ft-maps with r-mode injections, rows of $X'_2$  as points in a $d$-dimensional 
space. The idea behind the formulation of the SVM optimization problem is to find the optimal hypersurface that would separate (and hence classify) the noise points 
from the injection points. For this discussion we will need the following definitions: \\

\noindent
\textbf{Definition 1:} The distance of a point $x_i$ to a flat hypersurface $\mathcal{H} = \{ x | \langle w,x \rangle +b = 0 \}$ is given by
\begin{equation}
\label{eq.7a}
 d_{x_i}(w,b) = z_i \times (  \langle w,x_i \rangle +b )
\end{equation}

\noindent
where $w$ is a unit vector perpendicular to the flat hypersurface, $b$ is a constant,  and $z_i =+1$ for $\langle w,x_i \rangle +b >0$
and $z_i=-1$ $\langle w,x_i \rangle +b <0$. The index $i$ (in $x_i$) takes values from the set $\{1,2,3,..., 2N \}$. In the following discussion each point $x_i$ 
that lies above the hypersurface pairs with a value $z_i=1$ and each point $x_i$ that lies below the hypersurface pairs with a value of $z_i=-1$. \\

\noindent
\textbf{Definition 2:} The `margin', $\gamma_{\mathcal{S}}(w,b)$, of any set, $\mathcal{S}$, of vectors is defined as the minimum of the set of all 
distances $\mathcal{D} = \{d_{x_i}(w,b) | x_i \in \mathcal{S} \}$ from the hypersurface $\mathcal{H}$. For the purpose of our discussion the set $\mathcal{S}$ is the union of the
set of all noise points and the set of all injection points.\\

\noindent
For definition 3 we assume that a training set consists of points $x_i$ with each one belonging to one of two distinct data classes denoted by $y_i=1$ (for one class) 
and $y_i=-1$ (for the other class). We may further assume that the set of all noise points belongs to the class represented by $y_i=-1$ while the set of all injection points 
belongs to the class represented by $y_i=+1$.\\

\noindent
\textbf{Definition 3:} A training set $\{(x_1,y_1),...,(x_n,y_n) | x_i \in \mathbb{R}^d, y_i \in\{-1,+1 \} \}$ is called `separable'
by a hypersurface $\mathcal{H} = \{ x | \langle w,x \rangle +b =0 \}$ if both a unit vector $w$ $(\| w \|=1)$ and a constant $b$ exist so that the following inequalities hold:
\begin{align}
 \langle w,x_i \rangle +b & \ge \gamma_{\mathcal{S}} \,\,\,\,\,\   & \mbox{if} \,\,\,\,\,\   y_i=+1  \label{eq.8a} \\
 \langle w,x_i \rangle +b & \le -\gamma_{\mathcal{S}} \,\,\,\,\,\  & \mbox{if} \,\,\,\,\,\   y_i=-1  \label{eq.8b}
\end{align}

\noindent
where $\mathcal{S}=\{x_i | i=1,2,...,n \}$ and $\gamma_{\mathcal{S}}$ is given by definition 2. \\

For the purpose of our discussion $d$ is the dimensionality of the points $x_i$ (this dimensionality  corresponds to the number of pixels in each ft-map) and $n=2N$ is the 
number of our (ft-maps) data points. Using the fact that the hypersurface is defined up to a scaling factor $c$, i.e. $\mathcal{H} = \{ x | \langle cw,x \rangle +cb =0 \}$, we 
can take $c$ such that $c \gamma_{\mathcal{S}}=1$ and hence we can rewrite equations \eqref{eq.8a} and \eqref{eq.8b} as
\begin{equation}
\label{eq.9a}
 y_i \times ( \langle cw,x_i \rangle +cb ) \ge 1  \,\,\,\,\ \mbox{for all i=1,2,...,n}. 
\end{equation}

\noindent
Defining $w'=cw$ i.e. $\|w'\|=c$ and dividing equation \eqref{eq.9a} by $c$ we get
\begin{equation}
 y_i \times ( \langle \frac{w'}{\|w'\|},x_i \rangle + b ) \ge \frac{1}{\|w'\|}  \,\,\,\,\ \mbox{for all i=1,2,...,n}. 
\end{equation}

\noindent
\textbf{Formulation of the SVM optimization problem:}  Given a training set, that is, a data matrix 
$X' = \left( \begin{array}{c}
X'_1\\
X'_2 
\end{array} \right) $, $X'_1$ being a $N \times d$ matrix representing the noise points and $X'_2$ being a $N \times d$ matrix representing the injection points, 
we want to find the `optimal separating hypersurface' (OSH), that separates the row-vectors of $X'_1$ from the row-vectors of $X'_2$. According to definition 3, this translates to 
maximizing the `margin' $\gamma_{\mathcal{S}}$. In other words, we want to find a unit vector $w$ and a constant $b$ that maximize $\frac{1}{\|w'\|}$. Therefore, the SVM optimization 
problem can be expressed as follows
\begin{align}
 \min_{w,b} \,\,\  & \frac{1}{2} \|w'\|^2 \,\,\,\,\,\,\  \mbox{subject to}  \label{eq.10a} \\
  1-y_i \times ( \langle w',x_i \rangle & + b') \le 0  \,\,\,\,\ \mbox{for all i=1,2,...,n}  \label{eq.10b}
\end{align}

\noindent
where $b'=cb$. This is a quadratic (convex) optimization problem with linear constraints and can be solved by seeking a solution to the 
Lagrangian problem dual to equations \eqref{eq.10a} and \eqref{eq.10b}. 
  
Before formulating the Lagrangian dual we introduce the `slack variables', $\xi_i$ ($i=1,2,...,n$), that are used to relax the conditions in equation \eqref{eq.9a} and account for 
outliers or `errors'. Instead of solving equation \eqref{eq.10a} we seek a solution to 
 \begin{equation}
  \label{eq.11a}
 \begin{split}
 \min_{w,b} \,\,\  & \frac{1}{2} \|w'\|^2 + C \sum_{i=1}^n \xi_i \,\,\,\,\,\,\  \mbox{subject to}  \\
  \xi_i \ge 0 \,\,\ \mbox{and} \,\,\ 1-y_i \times ( & \langle w',x_i \rangle  + b')-\xi_i \le 0 \,\,\ \mbox{for all i=1,..,n}.
\end{split}
\end{equation}
 
 \noindent
 The slack variables $\xi_i$ measure the distance of a point that lies on the wrong side of its `margin hypersurface'. 
 Using the Lagrange multipliers 
 \begin{equation}
 \label{eq.mult}
  \alpha_i \ge 0 \,\,\,\  \text{and} \,\,\,\  \beta_i \ge 0
 \end{equation}

 \noindent 
 the Lagrangian dual formulation of equation \eqref{eq.11a} is to maximize the following Lagrangian
\begin{equation}
\label{eq.13a}
 \begin{split}
  \mathcal{L}(w^\prime,b,\xi_i,\alpha,\beta)= &  \frac{1}{2} \|w^\prime \|^2 + C \sum_{i=1}^n \xi_i  - \sum_{i=1}^n \beta_i \xi_i +   \\
  & +\sum^n_{i=1} \alpha_i(1-y_i \times ( \langle w^\prime,x_i \rangle +b) -\xi_i).
 \end{split}
\end{equation}

 \noindent
 Using the stationary first order conditions for $w^\prime$, $b$ and $\xi_i$ 
\begin{subequations} \label{eq:Derivatives}
\begin{align}
\frac{\partial  \mathcal{L}}{\partial w_j^\prime} &=  {w_j^\prime} -  \sum_{i=1}^n \alpha_i y_i x_{ij} = 0, \,\,\,\,\ \ \forall j=1,2, \dots d, & \label{eq.14b_1} \\ 
\frac{\partial  \mathcal{L}}{\partial b} &= \sum_{i=1}^n  \alpha_i  y_i  =  0,  & \\ \ 
\frac{\partial  \mathcal{L}}{\partial \xi_i} &= C - \alpha_i - \beta_i = 0, \,\,\,\,\ \ \forall i=1,2, \dots n \label{eq.14b_3}. 
\end{align}
\end{subequations} 

\noindent
(where $x_{ij}$ is the $j^{th}$ entry of the $x_i$ data point) the Lagrangian dual as given in expression \eqref{eq.13a} can be re-expressed only in terms of 
the $\alpha_i$ Lagrange multipliers, as follows
\begin{equation}
\label{Lag_dual}
 \mathcal{L}(\alpha_i) = \sum_{i=1}^n \alpha_i - \frac{1}{2} \sum_{i,j=1}^n \alpha_i \alpha_j y_i y_j \langle x_j, x_i \rangle
\end{equation}

\noindent
and hence we can evaluate the $\alpha_i$ Lagrange multipliers by solving the following optimization problem 
\begin{subequations} 
  \begin{align}
 \label{eqa:dual1}
 & \max_{\alpha_i} \mathcal{L}(\alpha_i)  \,\,\,\ \text{subject to} \,\,\,\,\ \sum_{i=1}^n \alpha_i  y_i  =  0, \\ 
 & 0 \leq \alpha_i \leq C \,\,\ \mbox{,} \,\,\ \forall i=1,2,.., n . \label{eqa:dual2}.
   \end{align}
\end{subequations}
 
 Defining $G_{ij}=y_i y_j x_j^\intercal x_i$ problem \eqref{eqa:dual1}-\eqref{eqa:dual2} is equivalently expressed as
 \begin{subequations}
 \label{eq.14b}
 \begin{align}
  & \min_{\alpha_i \in \mathcal{R}^n} \,\,\,\ \frac{1}{2} \alpha^\intercal G \alpha -e^\intercal \alpha \label{eq.14b1}\\
  & \mbox{subject to}  \,\,\ y^\intercal \alpha  = 0    \\
  & \mbox{and} \,\,\  0 \le \alpha_i \le C \,\,\ \mbox{,} \,\,\   \mbox{$i=1,2,...,n$} \label{eq.14b3}
  \end{align}
 \end{subequations}

 \noindent
 where $e^\intercal$ is a $n-$dimensional row vector equal to $e^\intercal=(1,1,...,1)$ and \eqref{eq.14b3} is derived from \eqref{eq.14b_3} together with \eqref{eq.mult}.
 
 Since the objective function in equation \eqref{eq.14b} is quadratic and all the constraints are affine, the problem defined by these equations is a quadratic 
optimization problem. Using the fact that (by constrution) the sum of all the entries of $G$ can be written as a sum of squares and also using that $\alpha_i \ge 0$ we can see
that $G$ is positive semidefinite, which implies that the problem is convex. Convex problems offer the advantage of global optimality; that is any local minimum is also the 
global one. Several methods have been proposed for solving such problems including primal, dual and parametric algorithms \citep{goldfarb1983numerically}.
 
 After solving the optimization problem defined by expressions \eqref{eq.14b1}-\eqref{eq.14b3}, i.e. after evaluating all the $\alpha_i$ ($i=1,2,...,n$), we can find the vector 
 $w$ using \eqref{eq.14b_1}. The constant $b$ can be found by using the Karush-Kuhn-Tucker (KKT) complementarity conditions \citep{fletcher2013practical}, 
 \begin{subequations} \label{eq.KKT}
 \begin{align}
  & \alpha_i\{-1+y_i \times (  \langle w',x_i \rangle  + b') + \xi_i \}= 0   \label{eq.KKT1} \\
  & \beta_i \xi_i = 0       \label{eq.KKT2}
  \end{align}
 \end{subequations}
 
 \noindent
along with equation \eqref{eq.14b_3}. For any $\alpha_i$ satisfying $0 < \alpha_i < C $, equation \eqref{eq.14b_3} implies that $\beta_i >0$ and hence \eqref{eq.KKT2} implies that $\xi_i = 0$. 
Consequently, we can use the $x_i$ corresponding to the aformentioned $\alpha_i$ to solve equation \eqref{eq.KKT1} for $b^\prime$. 

Having calculated the vector $w^\prime$ and the constant $b^\prime$ is equivalent to knowing the hypersurface defined by $\langle w',x_i \rangle  + b'=0$.
During the testing phase a new data point, $x_i$, is classified according to
\begin{equation}
\label{eq.testing}
\text{class}(x_i) = \text{sgn}(  \langle w',x_i \rangle  + b').
\end{equation}  

\noindent
For $\text{class}(x_i)=-1$ we classify the $x_i$ point as noise and for $\text{class}(x_i)=+1$ we classify the $x_i$ point as injection.

We choose to solve the convex quadratic problem as defined in equation \eqref{eq.14b} with sequential minimal optimization 
(SMO)\citep{platt1998sequential}. SMO modifies only a subset of dual variables $\alpha_i$ at each iteration,  and thus only some columns of $G$ are used at 
any one time. A smaller optimization subproblem is then solved, using the chosen subset of $\alpha_i$. In particular at each iteration only two Lagrange multipliers 
that can be optimized are computed. If a set of such multipliers cannot be found then the quadratic problem of size two is solved analytically. This process is 
repeated until convergence. The integrated software for support vector classification (LIBSVM) \citep{chang2011libsvm} is a state of the art SMO-type solver for the quadratic 
problem found in the SVM formulation. SMO outperforms most of the existing methods for solving quadratic problems \citep{platt1999fast}. Hence we choose to use it for 
training the SVM, using the LIBSVM routine `svmtrain'.
 
 \textbf{Non-linear SVM:} The soft margins $\xi_i$ can only help when data are `reasonably' linearly separable. However, in most real world problems, data is not linearly separable. 
 To deal with this issue we transform the data into a `feature' (Hilbert) space, $\mathcal{H}$, (a vector space equipped with a norm and an inner product), where a linear separation 
 might be possible due to the choice of the dimensionality of $\mathcal{H}$, $\text{dim}(\mathcal{H}) \ge \text{dim}(\mathbb{R}^d)$. The transformation is represented by 
 \begin{equation}
 \label{Phi}
 \begin{split}
   \Phi: & \mathbb{R}^d \rightarrow \mathcal{H} \\
  \mbox{such that} \,\ & \Phi(x_i) \in \mathcal{H}.
 \end{split}
 \end{equation}

 \noindent
 From equations \eqref{Lag_dual} and \eqref{Phi} we see that the non-linear SVM formulation depends on the data only through the dot products $\Phi(x_i) \cdot \Phi(x_j)$ in $\mathcal{H}$. 
 These dot products are generated by a real-valued `comparison function' (called the `Kernel' function) $k: \mathbb{R}^d \times \mathbb{R}^d \rightarrow \mathbb{R}$ that generates all the 
 pairwise comparisons $K_{ij}=k(x_i,x_j) =\Phi(x_i) \cdot \Phi(x_j)$. We represent the set of these pairwise similarities as entries in a $n \times n$ matrix, $K$. The use of a kernel function 
 implies that neither the feature transformation $\Phi$ nor the dimensionality of $\mathcal{H}$ are required to be explicitly known. \\
 
 \noindent
 \textbf{Definition 4:} A function $k: \mathcal{L} \times \mathcal{L} \rightarrow \mathbb{R}$ is called a positive semi-definite kernel if and only if it is:
 (i) symmetric, that is $k(x_i,x_j)=k(x_j,x_i)$ for any $ x_i, x_j \in \mathcal{L}$ and (ii) positive semi-definite, that is 
 \begin{equation}
 \label{eq.17a}
  c^\intercal Kc= \sum_{i=1}^n \sum_{j=1}^n c_i c_j k(x_i,x_j) \ge 0
 \end{equation}

 \noindent
 for any $x_i, x_j \in \mathcal{L}$ where $i,j \in \{1,2,...,n\}$ and any $c \in \mathbb{R}^n$ i.e. $c_i, c_j \in \mathbb{R} \,\,\ (i=1,2,...,n)$ and the $n \times n$ matrix $K$ has elements $K_{ij}=k(x_i,x_j)$.  \\
 
 The nature of the data we are using strongly suggests that our data points are not linearly separable in the original feature space. Therefore we choose to solve the dual 
 formulation as given by equation \eqref{eq.14b} where $G$ is now defined by $G_{ij}=y_i y_j k(x_i,x_j)$ so that we can use the `Kernel Trick'. Solving the dual problem has the additional advantage 
 of obtaining a sparse solution; most of the $\alpha_i$ will be zero (those that satisfy  $0< \alpha_i \le C$ are the support vectors that define the hypersurface). For the purpose of our study 
 we used the Radial Basis Function (RBF) kernel defined by 
 \begin{equation}
 \label{eq.18b}
  k(x_i,x_j)= \exp \Bigg ( - \gamma \frac{\|x_i-x_j  \|^2}{\sigma^2}   \Bigg ) 
 \end{equation}

\noindent
where $\gamma$ is a free parameter and $\sigma$ is the standard deviation of the $x_i$ that is equal to 1 due to normalization. Typically the free parameters ($\gamma$ and $C$) are calculated by using the cross validation 
(grid search) method on the data set, meaning that we split the data set into several subsets and the optimization problem is solved on each subset with different parameter values for $\gamma$ and $C$. 
We then choose the parameter values that give the lowest minimum value of the objective function. In our study we chose the default (by libsvm) value of $\gamma$ that was set equal to $\gamma = 1/d$. 
To determine the value of the parameter $C$, we plotted training efficiencies against several values of $C$. We determined that $C$ should be in the range of $10^4 - 10^5$. All experiments with SVM are conducted with 90/10 
split on data, where 90\% of the data is randomly selected for training and the remaining 10\% is used for testing. 
 
 Using the 'Kernel trick', we substitute $x_i$ with $\Phi(x_i)$ in equations \eqref{eq.11a}-\eqref{eq.testing}. Then equation \eqref{Lag_dual} is re-expressed as
 \begin{equation}
  \mathcal{L}(\alpha_i) = \sum_{i=1}^n \alpha_i - \frac{1}{2} \sum_{i,j=1}^n \alpha_i \alpha_j y_i y_j \langle \Phi(x_j), \Phi(x_i) \rangle.
 \end{equation}

 \noindent
 After solving \eqref{eq.14b}, the $\alpha_i$ ($i=1,2,...,n$) are substituted in \eqref{eq.14b_1} that we solve for $w_j^\prime$ to get
 \begin{equation}
 \label{w_prime}
  w_j^\prime =  \sum_{i=1}^n \alpha_i y_i \Phi_j(x_i) \,\,\,\,\ \ \forall j=1,2, \dots d 
 \end{equation}

 \noindent
 where $\Phi_j(x_i)$ is the $j^{th}$ entry of the $\Phi(x_i)$ transformed data point. Since the transformation $\Phi$ is not obtained directly we never calculate the $w^\prime$ vector explicitly.
 Nevertheless,we can substitute expression \eqref{w_prime} in \eqref{eq.KKT1} and solve the latter for $b^\prime$ (when $\xi_k=0$ and $\alpha_k \ne 0$) as follows
 \begin{equation}
 \label{b_prime}
  b^\prime = 1 - y_k \times  \sum_{i=1}^n \alpha_i y_i \langle \Phi(x_i) , \Phi(x_k) \rangle   
 \end{equation}

 \noindent
 where this result should be independent of which $k$ we use. Having the expression \eqref{w_prime} for the vector $w^\prime$ and the expression \eqref{b_prime} for the constant $b^\prime$ we can 
 classify a new data point during the testing phase according to
\begin{equation}
\label{eq.testing2}
\text{class}(x_i) = \text{sgn}(  \langle w',\Phi(x_i) \rangle  + b').
\end{equation}  
 
 \noindent
 From \eqref{eq.testing2} we see that we are able to calculate the new (flat) hypersurface in the new feature (Hilbert) space simply through inner products of 
 $\langle \Phi(x_i), \Phi(x_j)  \rangle $.

\section{Constrained Subspace Classifier}
\label{apCSC}

The idea in the constrained subspace classifier (CSC) method is similar to the idea used in SVM. In the latter the target was to separate the noise points 
(or noise vectors) from the injection points (or injection vectors) using a hypersurface. In the CSC method the idea is to project the noise vectors, 
rows of $X'_1$ ($N \times d$ matrix), onto a $d_1$-dimensional subspace $S_1$, (of dimensionality $d_1 < d$) of the $d$-dimensional space and also project 
the injection vectors, rows of $X'_2$ (also a $N \times d$ matrix), onto a subspace $S_2$, (of dimensionality $d_2 < d$). That is we seek to find two (optimal) 
subspaces such that we can classify data (ft-map) points according to their distance from each subspace: points closer to the subspace $S_1$ are classified as 
`noise points' and points closer to the subspace $S_2$ are classified as injection points. 

The optimality of the choice of each subspace depends on the chosen basis vectors, the chosen dimensionalities, $d_1$ and $d_2$, of each subspace as well as the
relative orientation between the two subspaces. Each choice corresponds to a given variance of the projected data: the closer the variance of the projected points 
is to the variance of the original data set the more optimal the subspaces are considered.

\subsection{The projection operator}

Let $S$ be a data space of dimension equal to the number of features, $d$, of the selected dataset (for our study $d$ is the dimensionality of the ft-maps 
after the resolution reduction). We can always find an orthonormal basis for $S$ (using the Gram-Schmidt process) given by
\begin{equation}
\label{eq.A1}
 U_d=\{u_1, u_2, \dots, u_d \} \,\,\ \mbox{with} 
 \,\,\ u_i \in \mathbb{R}^d \,\,\ \forall i=1,2,...,d
\end{equation}

\noindent
i.e. $U_d \in \mathbb{R}^{d \times d}$. We seek to find a subspace of $S$ of dimension $d_1 < d$. Since reducing the dimensionality brings the 
data points closer to each other, thus reducing the variance, we try to reduce the number of features from $d$ 
to $d_1$ while trying to maintain the variance of the data distribution as high as possible.  

To achieve the dimensionality reduction we seek to find a projection operator that projects the data points from $\mathbb{R}^{d}$ 
to a (dimensionally reduced) subspace $\mathbb{R}^{d_1}$ of orthonormal basis given by
\begin{equation}
\label{eq.A2}
U_{d_1}=\{ u_1, u_2, \dots, u_{d_1} \} \,\,\ \mbox{with} 
 \,\,\ u_i \in \mathbb{R}^d \,\,\ \forall i=1,2,...,d_1
\end{equation} 

\noindent
i.e. $U_{d_1}\in \mathbb{R}^{d \times d_1}$. By definition the projection operator is given by 
\begin{equation}
\label{eq.A3}
P=Q { ( Q^\intercal Q) }^{-1} Q^\intercal 
\end{equation}

\noindent
and projects a vector onto the space spanned by the columns of $Q$. Therefore, we may take the columns of $Q$ 
to be the orthonormal vectors given in \eqref{eq.A2}, that is $Q=U_{d_1}$. In that case, equation \eqref{eq.A3} becomes 
\begin{equation}
\label{eq.8}
P=U_{d_1} { (U_{d_1}^\intercal U_{d_1} ) }^{-1} U_{d_1}^\intercal 
\end{equation}

\noindent
which is the projection operator onto the space spanned by the column vectors of $U_{d_1}$. 

Since equation \eqref{eq.A1} is an orthonormal basis for $\mathbb{R}^d$ then $U_{d_1}^\intercal U_{d_1} = I_{d_1}$. Therefore, the expression of the projection 
operator that can project the (data) vectors in $\mathbb{R}^d$ onto its subspace $\mathbb{R}^{d_1}$ is given by 
\begin{equation}
\label{P}
 P=U_{d_1} U_{d_1}^\intercal.
\end{equation}

\noindent
In case $d_1=d$ then $P=U_{d} U_{d}^\intercal$. Since $U_d$ is a square matrix whose columns are orthonormal, this implies that its rows are also orthonormal. Orthonormality of the columns of $U_d$ implies 
$U_{d}^\intercal U_{d} = I_{d}$ (i.e. $U_{d}^\intercal$ is the left inverse of $U_d$) and orthonormality of the rows of $U_d$ implies $ U_d U_{d}^\intercal = I_{d}$
(i.e. $U_{d}^\intercal$ is the right inverse of $U_d$). Therefore, for the special case that $d_1=d$ we have that $U_{d}^\intercal$ is the inverse of $U_d$ or
\begin{equation}
\label{eq.inv}
 U_{d}^\intercal = U_{d}^{-1}.
\end{equation}

\subsection{Principal component analysis (PCA)}
\label{PCA}

To introduce PCA we will use the definition of the data matrix $X'_1$ as a $N \times d$ noise matrix as well as the definition of $X'_2$ as a $N \times d$ injection data matrix. 
Using the projection operator as given by expression \eqref{P} we want to project the ft-maps of $X'_1$ in a subspace $\mathbb{R}^{d_1}$ of $\mathbb{R}^d$ ($d_1 < d$). Let $x_i$ 
be the original $1 \times d$ row vector in $\mathbb{R}^d$. We project the column vector $x_i^\intercal$ onto $\mathbb{R}^{d_1}$ thus defining $\tilde{x_i}^\intercal = U_{d_1} U_{d_1}^\intercal x_i^\intercal$. 
Then the norm of the difference between the original and the projected (column) vectors can be expressed as 
\begin{equation}
 \| x_i^\intercal - \tilde{x_i}^\intercal \| = \| x_i^\intercal - U_{d_1} U_{d_1}^\intercal x_i^\intercal\|
\end{equation}
\noindent
where $U_{d_1} \in \mathbb{R}^{d \times d_1}$. In PCA we want to find the subspace $\mathbb{R}^{d_1}$ such that 
\begin{equation}
\begin{split}
\label{eq.12}
& \sum_{i=1}^{n} \| x_i^\intercal - U_{d_1} U_{d_1}^\intercal x_i^\intercal \|^2 \,\,\,\ \mbox{is minimized} \\
& \text{subject to} \,\,\,\,\ U_{d_1}^\intercal U_{d_1} = \mathcal{I}_{d_1}. \quad
\end{split}
\end{equation}

\noindent
This subspace $\mathbb{R}^{d_1}$ is defined as the $d_1$-dimensional hypersurface that is spanned by the (reduced) orthonormal basis 
$ \{ u_1, u_2, u_3, \dots, u_{d_1} \}$. i.e. finding such a basis is equivalent to defining the subspace $\mathbb{R}^{d_1}$.

\noindent
Using the definition of the Frobenius norm for a $m \times n$ matrix $A$, 
\begin{equation}
\label{eq.Frob}
 \| A \|_F = \sqrt{\sum_{i=1}^m \sum_{j=1}^n | a_{ij} |^2 } = \sqrt{ \mbox{trace} (A^* A)}
\end{equation}

\noindent
where $A^*$ is the conjugate transpose of $A$, we get 
\begin{equation}
\label{eq.10}
  \sum_{i=1}^{n}   \|  x_i^\intercal -  U_{d_1} U_{d_1}^ \intercal x_i^\intercal \|^2_{F} = \operatorname{tr} \big\{ {X'_1}^\intercal X'_1 ( \mathcal{I}-U_{d_1} U_{d_1}^\intercal ) \big\} 
\end{equation} 

\noindent
where $X'_1 \in \mathbb{R}^{n \times d}$ (where $n=2N$). Thus the optimization problem in equation \eqref{eq.12} reduces to \citep{vidal2005generalized}
\begin{equation}
\label{eq.11}
\begin{split}
&\min_{U_{d_1}} \operatorname{tr} \big \{  {X'_1}^\intercal X'_1 ( \mathcal{I} - U_{d_1} U_{d_1}^\intercal ) \big \} \\
& \text{subject to} \,\,\ U_{d_1}^\intercal U_{d_1} = \mathcal{I}_{d_1}. \quad 
\end{split}
\end{equation}

\noindent
Since  $ \operatorname{tr}\big\{ {X'_1}^{\intercal} X'_1 \big\} $ is a constant, the optimization problem can be re-written as
\begin{equation}
\label{eq.16}
\begin{split}
&\max_{U_{d_1}} \operatorname{tr}\{ U_{d_1}^{\intercal} {X'_1}^\intercal X'_1 U_{d_1} \} \\
& \text{subject to} \,\  U_{d_1}^\intercal U_{d_1} = \mathcal{I}_{d_1}. \quad \\
\end{split}
\end{equation}

To solve equation \eqref{eq.16} we define the Lagrangian dual problem by 
\begin{equation}
\begin{split}
\label{eq.17}
\mathcal{L} (U_{d_1}, \lambda_{ij} ) = & \operatorname{tr}( U_{d_1}^\intercal {X'_1}^\intercal X'_1 U_{d_1}) -  \\
 - & \sum_{i=1}^{d_1} \sum_{j=1}^{d_1} \lambda_{ij} ( \sum_{k=1}^d U_{jk}^\intercal U_{ki} - \delta_{ji} ) \\
& \text{where} \,\ \delta_{ij}= \left \{\begin{array}{ll} 
1 &  \text{for } i = j \\
0 &  \text{for } i \neq j .\\
\end{array}
\right.
\end{split}
\end{equation}

\noindent
Since $U_{d_1}^\intercal U_{d_1}$ is a symmetric $d_1 \times d_1$ matrix then the orthonormality condition in equation \eqref{eq.16} represents a total of
$d_1 \times (d_1 +1)/2$ conditions. Therefore, for the Lagrangian dual problem (as shown in equation \eqref{eq.17}) we need to introduce $d_1 \times (d_1 +1)/2$ 
Lagrange multipliers $\lambda_{ij}$. Hence we require that $\lambda_{ij}$ is a symmetric matrix. Also since each term in \eqref{eq.17} involves symmetric matrices
then the following first order optimality conditions
\begin{equation}
\label{eq.18}
\frac{\partial \mathcal{L}}{\partial \lambda_{pq}} = 0 \,\,\,\,\,\,\,\  \mbox{and} \,\,\,\,\,\,\,\ \frac{\partial \mathcal{L}}{\partial U_{lm}} = 0 .
\end{equation}

\noindent
can be solved for $\lambda_{ij}$ only if the latter is symmetric. Using equations \eqref{eq.17} and \eqref{eq.18} we get 
\begin{equation}
\label{eq.19}
\begin{aligned}
\frac{\partial}{\partial \lambda_{pq} } \bigg[ & \sum_{i=1}^{d_1} \sum_{j=1}^{d} \sum_{k=1}^{d} U_{ij}^\intercal (X^\intercal X)_{jk} U_{ki}  -  \\
 - & \sum_{i=1}^{d_1} \sum_{j=1}^{d_1} \lambda_{ij} ( \sum_{k=1}^d U_{jk}^\intercal U_{ki} - \delta_{ji} ) \bigg] = 0
\end{aligned}
\end{equation}

\noindent
and
\begin{equation}
\label{eq.21}
\begin{split}
\frac{\partial}{\partial U_{lm}} \bigg[ & \sum_{i=1}^{d_1} \sum_{j=1}^{d} \sum_{k=1}^{d} U_{ij}^\intercal (X^\intercal X)_{jk} U_{ki} -  \\
 - & \sum_{i=1}^{d_1} \sum_{j=1}^{d_1} \lambda_{ij} ( \sum_{k=1}^d U_{jk}^\intercal U_{ki} - \delta_{ji} ) \bigg] = 0. 
\end{split}
\end{equation}

\noindent
\noindent
Equation \eqref{eq.19} implies the $d_1 \times (d_1 +1)/2$ equations
\begin{equation}
\label{on}
 \sum_{k=1}^d U_{qk}^\intercal U_{kp} = \delta_{qp}
\end{equation}

\noindent
while equation \eqref{eq.21} implies the $d \times d_1$ equations
\begin{equation}
\label{eq.22}
\begin{split}
\sum_{j=1}^d U_{mj}^\intercal  ({X'_1}^\intercal X'_1)_{jl} + \sum_{k=1}^d  ({X'_1}^\intercal X'_1)_{lk} U_{km} - \\
- \sum_{j=1}^{d_1} \lambda_{mj} U_{jl}^\intercal  - \sum_{i=1}^{d_1} \lambda_{im} U_{li} = 0.
\end{split}
\end{equation}

\noindent
Using the fact that ${X'_1}^\intercal X'_1$ is symmetric, the first two terms of equation \eqref{eq.22} can be combined to a single term and similarly
(using the symmetry of $\lambda_{ij}$) the last two terms of equation \eqref{eq.22} can be combined to a single term to get
\begin{equation}
\label{eq.22a}
 \sum_{j=1}^d U_{mj}^\intercal  ({X'_1}^\intercal X'_1)_{jl} - \sum_{i=1}^{d_1} \lambda_{mi} U_{il}^\intercal = 0.
\end{equation}

\noindent
Equations \eqref{eq.22a} and \eqref{on} are sufficient to solve for $\lambda_{ij}$ and $U_{kl}$. Right-multiplying equation \eqref{eq.22a} by $U_{ln}$ and summing over $1 \le l \le d$ we get 
\begin{equation}
\label{eq.23b}
\sum_{l=1}^d \sum_{j=1}^d {U_{mj}}^\intercal  ({X'_1}^\intercal X'_1)_{jl} U_{ln} - \sum_{i=1}^{d_1} \lambda_{mi} \sum_{l=1}^d U_{il}^\intercal U_{ln} = 0.
\end{equation}

\noindent
Using equation \eqref{on} then equation \eqref{eq.23b} becomes 
\begin{equation}
\label{eq.24}
\sum_{l=1}^d \sum_{j=1}^d {U_{mj}}^\intercal  ({X'_1}^\intercal X'_1)_{jl} U_{ln} = \lambda_{mn}. 
\end{equation}

\noindent
Equations \eqref{eq.24} and \eqref{eq.22a} represent a set of $d_1 \times (d_1 +1)/2$ and $d_1 \times d$ equations respectively. These can be solved to obtain the 
$d_1 \times (d_1 +1)/2$ degrees of freedom of $\lambda_{ij}$ and the $d_1 \times d$ degrees of freedom of $U_{d_1}$.

The left hand side (LHS) of equation \eqref{eq.24} represents the $a_{mn}$ elements of a $d_1 \times d_1$ matrix and similarly the right hand side (RHS) of \eqref{eq.24} 
represents the $\lambda_{mn}$ elements of another $d_1 \times d_1$ matrix. Equation \eqref{eq.24} implies an entry-by-entry equation ($a_{mn}=\lambda_{mn}$) between the two matrices.  
Choosing $m=n$ and summing equation \eqref{eq.24} over $1 \le m \le d_1$ implies that the sum along the diagonal of the matrix on the LHS is equal to the sum
along the diagonal of the matrix on the RHS or equivalently  
\begin{equation}
\label{eq.27}
\sum_{m=1}^{d_1} \sum_{l=1}^d \sum_{j=1}^d {U_{mj}}^\intercal  ({X'_1}^\intercal X'_1)_{jl} U_{lm} = \sum_{m=1}^{d_1} \lambda_{mm}.
\end{equation}

\noindent
Noting that the LHS of \eqref{eq.27} is the trace of the LHS of \eqref{eq.24} we can re-write \eqref{eq.27} as
\begin{equation}
\label{eq.27a}
\operatorname{tr} (U_{d_1}^\intercal {X'_1}^\intercal X'_1 U_{d_1}) = \sum_{m=1}^{d_1} \lambda_{mm}.
\end{equation}

\noindent
To interpret the $\lambda_{mm}$ we use a theorem according to which the trace of a matrix is equal to the sum of its eigenvalues. Therefore, we can identify 
the $\lambda_{mm}$ for $1 \le m \le d_1$ as the eigenvalues of the symmetric matrix $(X_1 U_{d_1})^\intercal (X_1 U_{d_1})$. However, these $d_1$ eigenvalues 
are $d_1$ out of the total $d$ eigenvalues of $X_1^\intercal X_1$. This can be shown by using the invariance of trace under similarity transformations
(in this case under conjugacy). Using equation \eqref{eq.inv} we can re-write equation \eqref{eq.27a} for $d_1=d$ as
\begin{equation}
 \operatorname{tr} (U_{d}^{-1} {X'_1}^\intercal X'_1 U_{d}) = \operatorname{tr} ( {X'_1}^\intercal X'_1 ) = \sum_{m=1}^{d} \lambda_{mm}.
\end{equation}

\noindent
Therefore, the maximum of the objective function $F=\operatorname{tr} (U_{d_1}^\intercal {X'_1}^\intercal X'_1 U_{d_1})$ in expression \eqref{eq.16} is equal to the summation of the $d_1$ 
largest eigenvalues of $X_1^\intercal X_1$. Therefore the orthonormal basis for the lower dimensional subspace is given by the set of the eigenvectors corresponding 
to the $d_1$ largest eigenvalues of the symmetric matrix $X_1^\intercal X_1$.

\subsection{Formulation of CSC}

Consider the binary classification problem with $X'_1 \in \mathbb{R}^{n \times d}$ and $X'_2$ $\in \mathbb{R}^{n \times d}$ be the data matrices
corresponding to two data classes, $\mathcal{C}_1$ (noise points) and $\mathcal{C}_2$ (injection points) respectively. The number of data samples 
in $\mathcal{C}_1$ is the same as the number of data samples in $\mathcal{C}_2$ and is equal to $n/2$. The corresponding number of features is given 
by $d$ for both classes $\mathcal{C}_1$ and $\mathcal{C}_2$.  

We attempt to find two linear subspaces $\mathcal{S}_{1} \subseteq \mathcal{C}_1$ and $\mathcal{S}_{2} \subseteq \mathcal{C}_2$ 
that best approximate the data classes. Without loss of generality we assume the dimensionality of these subspaces to be the same and equal to $d_1$. Let 
\begin{equation}
\label{eq.28}
U = [u_1, u_2, \dots, u_{d_1}] \in \mathbb{R}^{d \times d_1}
\end{equation}

\noindent
and 
\begin{equation}
\label{eq.29}
V = [v_1, v_2, \dots, v_{d_1}] \in \mathbb{R}^{d \times d_1}
\end{equation}

\noindent
represent matrices whose columns are orthonormal bases of the subspaces $\mathcal{S}_{1}$ and $\mathcal{S}_{2}$ respectively. If we attempted to find 
$\mathcal{S}_{1}$ independently from $\mathcal{S}_{2}$ then we would have to capture the maximal variance of the data projected onto $\mathcal{S}_1$ 
separately from the maximal variance of the data projected onto $\mathcal{S}_2$. That would be equivalent to solving the following two optimization problems \citep{laaksonen1997local}
\begin{equation}
\label{eq.30}
\begin{split}
& \underset{U \in \mathbb{R}^{d \times d_1} } {\text{max}} \text{tr}( U^\intercal {X'_1}^\intercal X'_1 U) \\
& \text{subject to} \,\,\  U^\intercal U  = I_{d_1}
\end{split}
\end{equation}

\noindent
and
\begin{equation}
\label{eq.31}
\begin{split}
& \underset{V \in \mathbb{R}^{d \times d_1} } {\text{max}} \text{tr}( V^\intercal {X'_2}^\intercal X'_2 V) \\
& \text{subject to} \,\,\ V^\intercal V  = I_{d_1}.
\end{split}
\end{equation}

The solution to the optimization problem as shown in expression \eqref{eq.30} is given by the eigenvectors (the columns of the orthonormal basis $U$ of $\mathcal{S}_1$) 
corresponding to the $d_1$ largest eigenvalues of the matrix ${X'_1}^\intercal X'_1$. Similarly, the solution to the optimization problem as shown in expression \eqref{eq.31} 
is given by the eigenvectors (the columns of the orthonormal basis $V$ of $\mathcal{S}_2$) corresponding to the $d_1$ largest eigenvalues of the matrix ${X'_2}^\intercal X'_2$. 
Though the subspaces $\mathcal{S}_{1}$ and $\mathcal{S}_{2}$ are good approximations to the two classes  $\mathcal{C}_1$ and $\mathcal{C}_2$ respectively, these projections 
may not be the ideal ones for classification purposes as each one of them is obtained without the knowledge of the other. 

In the constrained subspace classifier (CSC) the two subspaces are found simultaneously by considering their relative orientation. This way CSC allows for a trade off between 
maximizing the variance of the projected data onto the two subspaces and the relative orientation between the two subspaces. The relative orientation between the two subspaces is generally defined 
in terms of the principal angles. The optimization problem in CSC is formulated as follows
\begin{equation}
\label{CSC}
\begin{split}
  \underset{ U, V \in \mathbb{R}^{d \times d_1}} {\text{max}} \text{tr}(U^\intercal {X'_1}^\intercal X'_1 U) & + \text{tr}( V^\intercal {X'_2}^\intercal  X'_2 V)+ \\
  & + C \text{tr} (U^\intercal  V V^\intercal U) \\
 \text{subject to} \quad U^\intercal U   = I_{d_1} , & \quad V^\intercal V  = I_{d_1}.
\end{split}
\end{equation}

\noindent
The last term of the objective function $G=\text{tr}(U^\intercal {X'_1}^\intercal X'_1 U)  + \text{tr}( V^\intercal {X'_2}^\intercal  X'_2 V)+ C \text{tr} (U^\intercal  V V^\intercal U)$ 
is a measure of the relative orientation between the two subspaces as defined in \citep{CSC_paper}. The parameter $C$ 
controls the trade off between the relative orientation of the subspaces and the cumulative variance of the data as projected onto the two subspaces. For large positive 
values of $C$, the relative orientation between the subspaces reduces (the two subspaces become more `parallel'), while for large negative values of $C$, the relative 
orientation increases (the two subspaces become more `perpendicular' to each other). 

This problem is solved using an alternating optimization algorithm described in \citep{CSC_paper}. For a fixed $V$, expression \eqref{CSC} reduces to
\begin{equation}
\label{CSC1}
\begin{aligned}
& \underset{ U \in \mathbb{R}^{d \times d_1}}{\text{max}} \text{tr}( U^\intercal ({X'_1}^\intercal X'_1 + C V V^\intercal) U) \\
& \text{subject to} \,\,\ U^\intercal U  = I_{d_1}.
\end{aligned}
\end{equation}

\noindent
The solution to the optimization problem \eqref{CSC1} is obtained by choosing the eigenvectors corresponding to the $d_1$ largest eigenvalues of the symmetric matrix 
${X'_1}^\intercal X_1 + C V V^\intercal$. Similarly, for a fixed $U$, expression \eqref{CSC} reduces to
\begin{equation}
\label{CSC2}
\begin{split}
& \underset{ V \in \mathbb{R}^{d \times d_1}} {\text{max}} \text{tr} ( V^\intercal ({X'_2}^\intercal X'_2 + C U U^\intercal) V) \\
& \text{subject to} \,\,\ V^\intercal V  = I_{d_1}
\end{split}
\end{equation}

\noindent
where the solution to the optimization problem \eqref{CSC2} is again obtained by choosing the eigenvectors corresponding to the $d_1$ largest eigenvalues of the symmetric 
matrix ${X'_2}^\intercal X'_2 + C U_1 U_1^\intercal$. 

The algorithm for CSC can be summarized as follows:
\begin{algorithm} [H]                   
\caption{CSC ($X'_1, X'_2$, $d_1$, $C$)}   
\label{alg1}                           
\begin{algorithmic}               
\begin{small}
\STATE 1. Initialize $U$ and $V$ such that $U^\intercal U  = I_{d_1}$, $V^\intercal V  = I_{d_1}$. 
\STATE 2. Find eigenvectors corresponding to the $d_1$ largest eigenvalues of the symmetric matrix $ {X'_1}^\intercal X'_1 + C V V^\intercal$. 
\STATE 3. Find eigenvectors corresponding to the $d_1$ largest eigenvalues of the symmetric matrix $ {X'_2}^\intercal X'_2 + C U U^\intercal$. 
\STATE 4. Alternate between 2 and 3 until one of the termination rules below is satisfied.
\end{small}
\end{algorithmic}
\end{algorithm}
  
\noindent  
We define the following three termination rules:
\begin{itemize}
\item Maximum limit $Z$ on the number of iterations, 
\item Relative change in $U$ and $V$ at iteration $m$ and $m+1$,
\begin{equation}
\label{eq.tol}
\begin{split}
\text{tol}_U^m =  & \frac{\| U^{(m+1)} - U^{(m)} \|_{F}}{\sqrt{N}}, \\
\text{tol}_V^m =  & \frac{\| V^{(m+1)} - V^{(m)} \|_{F}}{\sqrt{N}}
\end{split}
\end{equation}

\noindent
where $N$ = $d\times d_1$ and the subscript $F$ denotes the Frobenius norm. 
\item Relative change in the value of the objective function $G$ as shown in expression \eqref{CSC} at iteration $m$ and $m$+1,
\begin{equation}
\text{tol}_{f}^{m} = \frac{G^{(m+1)} - G^{(m)}}{|G^{(m)}| + 1}.
\end{equation}
\end{itemize}
 
\noindent
The value of Z was set to $2000$, while $\text{tol}_{f}^{m}$, $\text{tol}_U^m$ and $\text{tol}_V^m$ are all set at the same value of $10^{-6}$. From equation \eqref{eq.Frob} 
we see that the factor of $1/\sqrt{N}$ in \eqref{eq.tol} results in the averaging of the squares of all the entries of the matrices $(U^{(m+1)} - U^{(m)})$ or $(V^{(m+1)} - V^{(m)})$. 
This regularization factor keeps the tolerance values independent of the data set. \\
 
After solving the optimization problem \eqref{CSC} (by utilizing algorithm 2) a new point $x$ is classified by computing the distances from the two 
subspaces $\mathcal{S}_{1}$ and $\mathcal{S}_{2}$ defined by
\begin{equation}
\text{dist}( x,\mathcal{S}_1) = \text{tr}( U^\intercal x^\intercal x U)
\end{equation}

\noindent
and
\begin{equation}
\label{eq.32}
\text{dist}( x,\mathcal{S}_2) = \text{tr}( V^\intercal x^\intercal x V).
\end{equation}

\noindent
The class of $x$ is defined by
\begin{equation}
\label{eq.33}
\text{class}(x) = \text{arg} \{ \min_{ i \in \{1,2 \} } \{\text{dist}(x,\mathcal{S}_{i})\} \}.
\end{equation}

\noindent
In our case, if $x$ is closer to $S_1$ then $x$ is classified as noise (or `no signal') and if $x$ is closer to $S_2$ then $x$ is classified as 
an r-mode injection (or `presence of signal').

\section{Data Preparation}
\label{Dataprep}
  
\subsection{Production of the data matrix for the MLA training}

We start with the data maps in the frequency-time domain (ft-maps) produced by the stochastic transient analysis multi-detector pipeline (STAMP) \citep{STAMPPAPER}. Let $N$ 
be the number of noise maps. The number of (r-mode) injection maps is also equal to $N$. These ft-maps are produced using simulated data recolored with the aLIGO sensitivity 
noise curve. Each map has a size of $F \times T$ pixels with each pixel along the vertical axis corresponding to $\delta f$ Hz and each pixel along the horizontal axis corresponding 
to $\delta t$ s, hence the length of the map along the vertical axis is $(F \delta f)$ Hz and the length of the map along the horizontal axis is $(T \delta t)$ s. 
This ft-map is reshaped to a $1 \times D$ (where $D=F T$) row vector. We reshaped all $2N$ ft-maps (each one of size $F \times T$) and produced $2N$ row vectors $x_i$ 
with $i \in \{1,2,...,2N \}$. The rows with $i \in \{1,2,...,N \}$ correspond to the noise ft-maps while the rows with $i \in \{N+1,N+2,...,2N \}$ correspond to the injection ft-maps. 

We then used the rows $x_i$ with $i \in \{1,2,...,N \}$ to produce a $N \times D$ noise data matrix, $X_1$ and we also used the rows with $x_i$ with $i \in \{N+1,N+2,...,22700 \}$ 
to produce a $N \times D$ injection data matrix, $X_2$. The MLAs would take as an input the $2N \times D$ data matrix given by
\begin{equation}
\label{X}
X = \left( \begin{array}{c}
X_1\\
X_2 
\end{array} \right) . 
\end{equation} 

\noindent
Each row $x_i$ with $i \in \{1,2,..,2N \}$ of the data matrix $X$  corresponds to a single ft-map. The total number of rows is equal to the number of data points, $n=2N$, 
while the total number of columns (i.e. the total number of features) is equal to $D=F T$, where $D$ is the dimensionality of the feature space in which each single 
ft-map lives.  

For any matrix we know that row rank = column rank, therefore, the number of linearly independent columns of $X$ is equal to $2N$. This number is determined by the limited
number ($n=2N$) of ft-maps we could produce. This means that even though each single ft-map lives in a $D-$dimensional space ($D \gg n$), we can only approximate these ft-maps
as vectors living in a $n$-dimensional space (subspace of the $D-$dimensional space). The best approximation of this subspace would be the one in which the most 
'dominant' $n$ features (out of the total number of $D$) constitute a basis of the subspace. A well known method of choosing the $n$ most dominant features is described 
by the principal component analysis (PCA) \citep{jolliffe2002principal} or see section \ref{PCA}. However, the (RAM) memory required to perform PCA on $X$ is beyond $1$TB, 
thus making it practically impossible to perform PCA on $X$ with realistically available computing resources. 

A reliable approach to solve the problem of the high dimensionality of the features ($D \gg n$) is to seek MLAs that will naturally select $d$-many features (with $d \ll D$) 
such that $d \leq n$ \citep{johnstone2009statistical}. Three classes of MLAs that can achieve this are the ANN, SVM and CSC methods. However, the data matrix is too large to 
attempt to perform any MLAs on it. Therefore, the only way out of these restrictions the data matrix size imposes, is to perform resolution reduction for each $F \times T$ 
ft-map (before reshaping each one of them to a row vector). After the resolution reduction, performing further feature selection would still benefit the training of the algorithms 
in terms of speed. The right choice of features can significantly decrease the training time without noticeably affecting the training efficiencies.   

A resolution reduction on the ft-maps would result in a number of $2N$ row vectors (of dimensionality $1 \times d$) such that $d \ll D$. The desired effect of the resolution 
reduction would be to get $d \leq n$. The first guess for such a reduction would be to choose a factor of $D/n$. That would be equivalent to a reduction by a factor of 
$\sim \sqrt{\frac{D}{n}}$ along each axis (frequency and time) of the ft-map. However, it turned out that this is not the optimal resolution (per axis) reduction factor. 
The following two sub-sections describe the experimentation on the reduction factor.  

\subsection{Resolution reduction: bicubic interpolation}

To perform the resolution reduction, we used the imresize matlab function. The original ft-map of $F \times T$ pixels consists of a $(F+1) \times (T+1)$ point grid. Imresize will 
first decrease the number of points in the point grid according to the chosen resolution reduction factor, $r$. Interpolation is then used to calculate the surface within each pixel 
in the new point grid. The result is a new ft-map of dimensionality $\frac{F}{r} \times \frac{T}{r}$ with a number of pixels equal to $d=\frac{FT}{r^2}=\frac{D}{r^2}$. 

We used the bicubic interpolation option of the imresize function. According to this, the surface within each pixel can be expressed by
\begin{equation}
 S(t,f)= \sum^3_{i=0}  \sum^3_{j=0} a_{ij} t^i f^j  
\end{equation}

\noindent
The bicubic interpolation problem is to calculate the $16$ $a_{ij}$ coefficients. The $16$ equations used for these calculations consist of the following conditions at the 4 
corners of each pixel: \\
(a) the values of $S(t,f)$  \\
(b) the derivatives of $S(t,f)$ with respect to $t$ \\
(c) the derivatives of $S(t,f)$ with respect to $f$ and \\
(d) the cross derivatives of $S(t,f)$ with respect to $t$ and $f$ \\

Determining the resolution reduction factor that would yield the best training efficiencies for the MLAs was not a very straight forward task. To do so we performed a series 
of tests using the set of $N$ noise ft-maps and the set of $N$ injection ft-maps. The injected signal SNR values lay in a range such that $10^{-23.7} \le h \le 10^{-23.2}$.  

\subsection{Resolution reduction versus training efficiency}

\begin{figure}
\includegraphics[width=1.05 \linewidth]{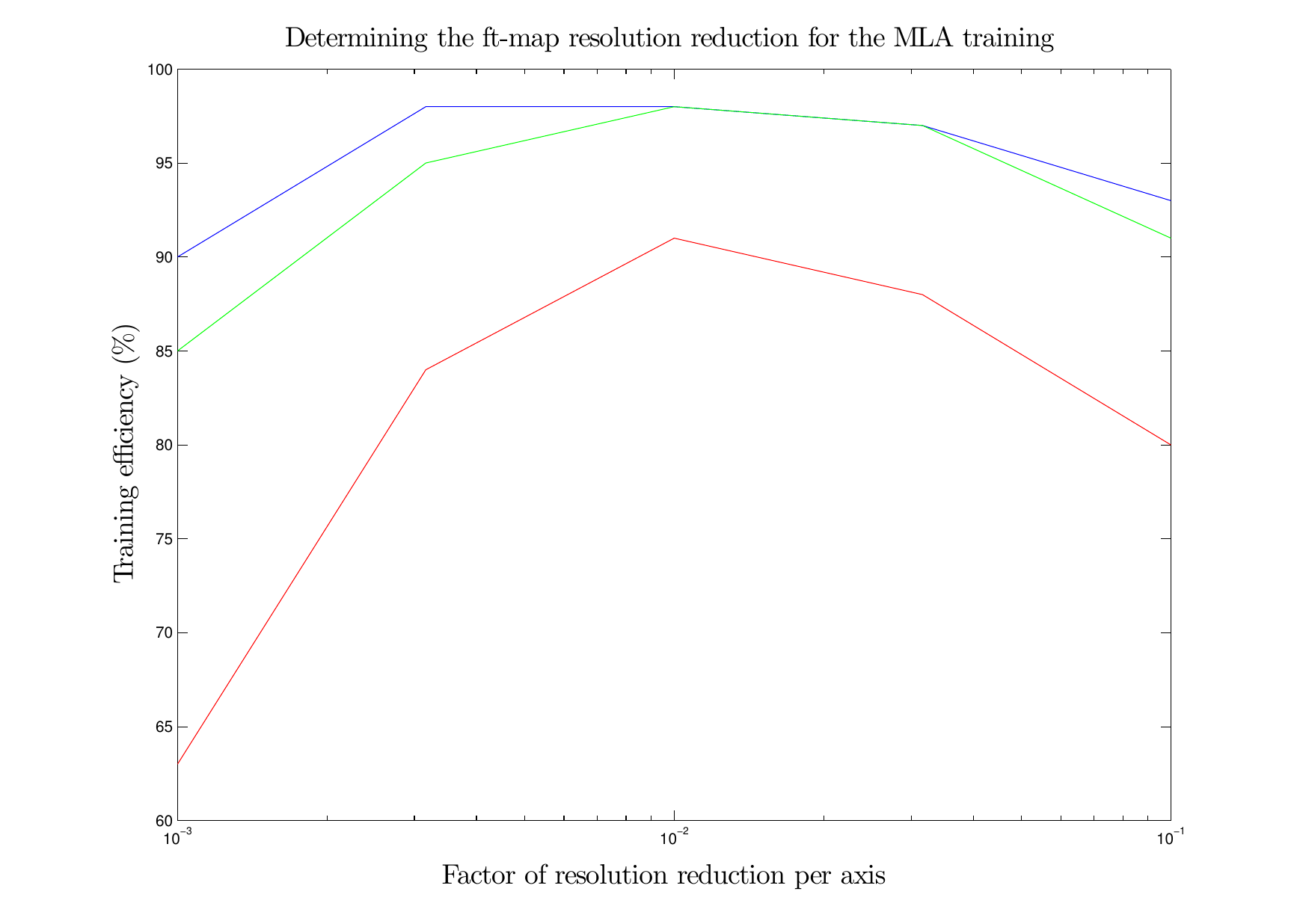}
\caption{Training efficiencies of ANN (blue), SVM (green) and CSC (red) versus the resolution reduction (per axis). For SVM and CSC there is a clear peak 
at a resolution reduction factor of $10^{-2}$. The ANN peak seems to be a little off but for uniformity we used $10^{-2}$ for all 3 MLAs. 
The training of all three MLAs was performed using the ($\alpha=0.1, f_o=1500$) waveform. No tests have been performed to verify the validity of these plots for 
other waveforms or other $h$ value ranges.}  \label{Fig:ef11}
\end{figure}

We tested 5 different resolution reduction factors ($r=10^{-1}$, $r=10^{-1.5}$, $r=10^{-2}$, $r=10^{-2.5}$ and $r=10^{-3}$) where the value of $r$ corresponds to the factor
by which each axis resolution is reduced. With $N=11350$ and $F=1001$, $T=4999$ (such that $D=5003999$) the resulting ($2N \times d$) data matrices had dimensions $22700 \times 50500$, 
$22700 \times 5155$, $22700 \times 550$, $22700 \times 64$ and $22700 \times 10$ respectively. Subsequently each of the three MLAs were trained and the training efficiencies 
were plotted against the resolution reduction factors. The results are shown in Fig.\ref{Fig:ef11}. From the plots we see that the training efficiencies first improve as we 
lower the resolution. For too low or too high resolution reductions the training efficiencies decrease. This behavior was consistent on all three MLAs. At a reduction factor 
of 100 per axis we have the maximum training efficiency. Resolution reduction offers two advantages: (a) it increases the MLA training efficiency and (b) it reduces the training 
time. Using the results from Fig.\ref{Fig:ef11} we determined that the best resolution reduction would be the factor of $r=10^{-2}$. This results in a data matrix with dimensions 
of $22700 \times 550$ (disc space of 84MB).

After dimensionality reduction, the matrices $X_1$ and $X_2$ become $X'_1$ with row vectors $x_i \in \mathbb{R}^{550}$ where $i \in \{1,2,...,11350 \}$ and $X'_2$ with row 
vectors $x_i \in \mathbb{R}^{550}$ where $i \in \{11351,...,22700 \}$. Both of the $X'_1$ and $X'_2$ has a reduced dimensionality $11350 \times 550$. Similarly we define the 
dimensionally reduced $22700 \times 550$ data matrix 
\begin{equation}
\label{X'}
X' = \left( \begin{array}{c}
X'_1\\
X'_2 
\end{array} \right). 
\end{equation} 

\noindent
The number of rows, $n=22700$, is the number of data points (ft-maps) and the number of columns, $d=550$, is the number of features of 
each point or the dimensionality of the space in which each ft-map lives (after the resolution reduction).

%\bibliography{r_modes_bib.bib}

\end{document}